\documentclass[a4paper,11pt]{article}
\pdfoutput=1
\usepackage[dvipsnames]{xcolor}
\usepackage{jcappub}
\usepackage[T1]{fontenc}
\usepackage{amsmath}
\usepackage{amssymb}
\usepackage{graphicx}
\usepackage{booktabs}
\usepackage{stackengine}
\usepackage{mathtools}

\usepackage{soul}
\stackMath
\usepackage{array}
\usepackage[utf8]{inputenc}
\usepackage{geometry}
\usepackage{svg}

\usepackage{subfig}
\usepackage{placeins}
\usepackage{relsize}
\setul{5pt}{1.5pt}

\def\mathclap#1{\text{\hbox to 0pt{\hss$\mathsurround=0pt#1$\hss}}}
\def\bclap#1{\,\text{\hbox to 0pt{\hss$\mathsurround=0pt#1$\hss}}\,}

\newcommand{\lcdm}{$\Lambda$CDM}
\newcommand{\bessel}[2]{\skew{5}{\dot}\jmath_{#1}(#2)}

\newcommand{\dnew}{\newline \newline}

\newcommand{\kmin}{\ensuremath{k_\mathrm{min}}}
\newcommand{\kmax}{\ensuremath{k_\mathrm{max}}}
\newcommand{\hyp}[4]{\ensuremath{{}_2F_1\left(#1,#2;#3\enskip \vline \enskip#4\right)}}
\newcommand{\sumn}{\ensuremath{\sum\limits_n}}
\newcommand{\summ}{\ensuremath{\sum\limits_m}}
\newcommand{\sumx}{\ensuremath{\sum\limits_x}}

\newcommand{\sumxy}{\ensuremath{\sum\limits_{x,y}}}
\newcommand{\tmin}{\ensuremath{t_\mathrm{min}}}

\newcommand{\xmin}{\ensuremath{\chi_\mathrm{min}}}

\newcommand{\xmax}{\ensuremath{\chi_\mathrm{max}}}
\newcommand\Tstrut[1]{\rule{0pt}{#1ex}}       
\newcommand{\rsd}{Redshift-Space Distortion}
\newcommand{\tquote}[1]{``#1''}

\newcommand{\n}{\hat{\boldsymbol n}}
\newcommand{\vk}{\boldsymbol k}
\newcommand{\vx}{\boldsymbol x}
\newcommand{\intinf}{\int_0^\infty}

\def\mylist#1 {\ifx\endmylist#1\else\makebox[4em][r]{#1} \expandafter\mylist\fi}
\def\endmylist{}
\title{
	Beyond the traditional Line-of-Sight approach of cosmological angular statistics
	}
\author[a]{Nils Sch\"oneberg}
\author[b,c]{Marko Simonovi\'c}
\author[a]{Julien Lesgourgues}
\author[b]{Matias Zaldarriaga}
\affiliation[a]{RWTH Aachen University, \textit{Institute for Theoretical Particle Physics and Cosmology (TTK)}, Aachen, Germany}
\affiliation[b]{School of Natural Sciences,\textit{ Institute for Advanced Study}, \\1 Einstein Dr, Princeton, 08540 NJ, USA}
\affiliation[c]{Theoretical Physics Department, CERN, 1211 Geneva 23, Switzerland}
\emailAdd{schoeneberg@physik.rwth-aachen.de}
\emailAdd{lesgourg@physik.rwth-aachen.de}
\emailAdd{marko.simonovic@cern.ch}
\emailAdd{matiasz@ias.edu}
\abstract{We present a new efficient method to compute the angular power spectra of large-scale structure observables that circumvents the numerical integration over Bessel functions, expanding on a recently proposed algorithm based on FFTlog. This new approach has better convergence properties. 
The method is explicitly implemented in the \texttt{CLASS} code for the case of number count $C_\ell$'s (including redshift-space distortions, weak lensing, and all other relativistic corrections) and cosmic shear $C_\ell$'s. In both cases our approach speeds up the calculation of the exact $C_\ell$'s (without the Limber approximation) by a factor of order 400 at a fixed precision target of 0.1\%. 
}
\date{\today}
\begin{document}

	\hfill{\small TTK-18-28}

	\maketitle
	\flushbottom
	\section{Introduction}
One important pathway to constrain cosmological parameters is the measurement of the two-point statistics of large-scale structure. For galaxy redshift surveys, this measurement is often cast in a three-dimensional redshift-dependent Fourier power spectrum $P(k,z)$. However, this quantity is not directly observable, because redshift surveys do not map objects in four-dimensional spacetime, but rather in the three-dimensional past-light-cone, with one redshift and two angular coordinates. Thus $P(k,z)$ can only be reconstructed by assuming a fiducial cosmology and by going through some rather complicated steps. A more natural and straightforward way to describe the two-point statistics of a galaxy catalogue is through the angular number count power spectra $C_\ell$'s for a list of redshift bins, including auto- and cross-correlations between bins. The bin width usually reflects the  measurement error of the (photometric or spectroscopic) survey. 
\dnew
Although the angular spectra are a much more direct representation of the data, analysis pipelines based on the matter power spectrum are often preferred for computational reasons. Indeed, an accurate calculation of the angular spectra in many redshift bins (beyond the Limber approximation) is very time-consuming compared to the calculation of the matter power spectrum. This problem is worsened when all the General Relativity (GR) corrections to the number count $C_\ell$'s, presented in \cite{Yoo:2009au,Yoo:2010ni,Bonvin:2011bg,Challinor:2011bk,Bertacca2012tp,DiDio:2013bqa} and implemented in the codes \texttt{CAMB\_sources}\footnote[1]{https://camb.info/sources/} \cite{Lewis:1999bs,Challinor:2011bk} and \texttt{CLASS}\footnote[2]{http://class-code.net/} \cite{Blas:2011rf,DiDio:2013bqa}, are taken into account. Then, even some idealized Fisher matrix forecasts can become computationally expensive \cite{DiDio:2013sea,DiDio:2016ykq}, and even more so MCMC parameter extractions. While some of the GR corrections can always be safely neglected, the \rsd s (RSDs) will have to be taken into account in the analysis of future high-precision galaxy redshift surveys. The weak lensing contributions can also be significant, especially in the case of cross-correlation spectra between redshift bins \cite{DiDio:2013sea}.
\dnew
For weak lensing (i.e. cosmic shear) surveys, the data is more often expressed in terms of angular power spectra $C_\ell$'s in a list of redshift bins, with an important role of the cross-correlation spectra even for non-adjacent bins (since weak lensing effects are correlated for sources laying at different redshifts). The Limber approximation based on the matter power spectrum $P(k,z)$ is often used at high $\ell$'s in order to speed up the calculation, but the low-$\ell$ part of the calculation remains slow, and the matching between the exact spectrum and the Limber spectrum at some intermediate $\ell$ value is a source of inaccuracy.
\dnew
For both redshift and cosmic shear surveys, there exist some alternative ways to represent the two-point statistics of the data, such as two-point correlation functions in redshift shells (see \cite{Tansella:2017rpi}). An algorithm efficiently computing these functions has recently been described in \cite{Tansella:2018sld}.
The purpose of this paper is, however, to stick to the angular spectra and to re-visit the numerical method used for calculating $C_\ell$'s, with a new way to separate the cosmological and geometrical information, even more efficient than the traditional line-of-sight approach of \cite{Seljak:1996is}. The main ideas of this method were suggested recently in \cite{Assassi:2017lea,Gebhardt:2017chz}, and are based on a power law decomposition of the source functions, which allows to separate the integrals into one part depending on cosmology and one part depending on geometry. The integration of spherical Bessel functions, which are quickly oscillating and only slowly damped, can then be performed analytically. The new method speeds up the calculation of the number count and cosmic shear $C_\ell$'s by one to two orders of magnitude (when weak lensing corrections are taken into account, and also for number count). Thus our work contains a step towards practical implementations of angular power spectra in the analysis pipeline of future galaxy redshift surveys and cosmic shear surveys.\pagebreak[3]
\dnew
In section \ref{sec_method}, we explain the general approach, including new developments with respect to reference~\cite{Assassi:2017lea}. In section \ref{sec_types} we discuss our practical implementation of this method in \texttt{ CLASS}, including all GR corrections to the number count spectra $C_\ell$'s, and introducing the idea of \tquote{approximate separability}.  Finally, in section \ref{sec_concl}, we give a report on the performance of the new code (for a fixed level of precision) and conclude. The reader is encouraged to look at \cite{Assassi:2017lea} and \cite{Gebhardt:2017chz} for more theoretical and technical details on this method and its application to other types of spectra (such as the CMB power spectrum or general bispectra).

	\section{Method}\label{sec_method}

\subsection{The Traditional Line-of-sight Approach}	
	
Any observable $\mathcal{O}_\alpha(\n,z)$ can be expanded for every redshift $z$ in spherical harmonics $Y_{\ell m}(\n)$ with respect to its angular position on the sky $\n$,
\begin{equation}
\mathcal{O}_\alpha(\n, z) = \sum\limits_{\ell,m} a^\alpha_{\ell m}(z)Y_{\ell m}(\n)\;, \qquad \text{where} \quad \enskip a^\alpha_{\ell m}(z) = \int d\Omega_{\n} Y_{\ell m}^{*}(\n) \mathcal{O}_\alpha(\n, z) ~.
\end{equation}
The angular expansion coefficients $a^\alpha_{\ell m}$ can be used to define the angular power spectrum for the observables $\mathcal{O}_\alpha$, or more generally any cross-spectrum
\begin{equation} \label{eq_power_theory}
C_\ell^{\alpha \beta}(z_1,z_2) \equiv \left\langle a^\alpha_{\ell m}(z_1) \,\, { a^\beta_{\ell m}(z_2)}^{*} \right\rangle ~.
\end{equation}
It is well-known that the theoretical prediction for the coefficients $a^\alpha_{\ell m}(z)$ is given by a convolution of the Fourier transform of the observable in real space $\mathcal{O}_\alpha(\vk, z)$ with a spherical Bessel function $\bessel{\ell}{k \chi(z)}$,
\begin{equation} \label{eq_angular_theory}
a^\alpha_{\ell m}(z) = 4 \pi i^\ell \int \frac{d^3k}{(2 \pi)^3} \, \bessel{\ell}{k \chi(z)} \, \mathcal{O}_\alpha(\vk, z) \, Y_{\ell m}^{*}(\hat \vk) ~.
\end{equation}
Here $\vk$ is a Fourier wave number, $k = | \vk |$ its modulus, $\hat \vk = \vk / k$ the corresponding unit vector, $\chi(z)$ is the comoving distance, given by
\begin{equation}
\chi(z) = \int_0^z \frac{dz'}{a_0 H(z')}~, 
\end{equation}
and $H(z)$ is the Hubble parameter. In the following we will neglect the explicit dependence and write $\chi \equiv \chi(z)$. The quantity $\mathcal{O}_\alpha(\vk,z)$ is usually computed by Einstein-Boltzmann solvers like \texttt{CAMB} or \texttt{CLASS}. It can be split into a transfer function $T_\alpha(k,z)$ and a primordial curvature perturbation $\mathcal{R}(\vk)$,
\begin{equation} \label{eq_obs_separation_trans_prim}
\mathcal{O}_\alpha(\vk,z) = T_\alpha(k,z) \mathcal{R}(\vk)~.
\end{equation} 
The primordial curvature perturbation is randomly distributed according to a (nearly scale-independent) primordial spectrum $\mathcal{P}_\mathcal{R}(k)$, defined in the following way
\begin{equation}
\langle \mathcal{R}(\vk)\mathcal{R}(\vk') \rangle \equiv \frac{2\pi^2}{k^3} \mathcal{P}_\mathcal{R}(k) \, \delta(\vk - \vk') \;. 
\end{equation}
Plugging (\ref{eq_obs_separation_trans_prim}) into (\ref{eq_angular_theory}) and finally into (\ref{eq_power_theory}), we obtain
\begin{equation} \label{eq_power_final}
C_\ell^{\alpha \beta}(z_1,z_2) = 4 \pi \intinf \frac{dk}{k} \, \, \mathcal{P}_\mathcal{R}(k) \,\, T_\alpha(k,z_1) \bessel{\ell}{k \chi_1}\,\, T_\beta(k,z_2) \bessel{\ell}{k \chi_2}~.
\end{equation}
Furthermore, in concrete observables like galaxy number count or cosmic shear, redshift measurement errors are described by averaging $\mathcal{O}_\alpha(\n, z)$ over some window functions $\mathcal{W}^i(z)$ normalised to one ($i$ is the index of a given redshift bin).
The normalised window function in comoving distance space is then $W^i(\chi) = \mathcal{W}^i(z) a_0 H(z)$, where $z$ is understood as a function of $\chi$, such that
\begin{equation}
\intinf W^i(\chi) d\chi = \intinf \mathcal{W}^i(z) a_0 H(z)d\chi = \intinf \mathcal{W}^i(z)dz = 1~.
\end{equation}
Averaging equation (\ref{eq_power_final}) over these windows leads to the expression 
\begin{align}
C_\ell^{\alpha\beta,ij} = 4 \pi & \intinf d\chi_1\, W^i(\chi_1) \, \intinf d\chi_2 \, W^j(\chi_2) \nonumber \\
&\qquad  \times \intinf \, \frac{dk}{k} \, \, \mathcal{P}_\mathcal{R}(k) \,\, T_\alpha(k,\chi_1) \bessel{\ell}{k \chi_1}\,\, T_\beta(k,\chi_2) \bessel{\ell}{k \chi_2}~.
\end{align}
The usual way to solve such an integral is to calculate  
\begin{equation} \label{eq_old}
C_\ell^{\alpha\beta,ij} =  4 \pi \intinf \frac{dk}{k} \, \mathcal{P}_\mathcal{R}(k) \, \Delta^{\alpha,i}_\ell(k)\, \Delta^{\beta,j}_\ell(k) ~,
\end{equation}
where for every wavenumber $k$ and every window function $i$ the integral 
\begin{equation} \label{eq_traditional_los}
\Delta^{\alpha,i}_\ell(k) = \intinf d\chi \, W^i(\chi) T_\alpha(k,\chi) \bessel{\ell}{k \chi}
\end{equation}
has to be computed. The Spherical Bessel functions $j_\ell(x)$ introduce fast and slowly damped oscillations, making it hard to compute the integrals efficiently. 
\subsection{Separating Cosmology and Geometry}\label{sec_cosm_geom}
It has been argued in \cite{Assassi:2017lea} that there is a simple way to circumvent the issue with the integrals over spherical Bessel functions. The main idea is to approximate the $k$-dependence of the product $\mathcal{P}_\mathcal{R}(k) T_\alpha(k,\chi_1) T_\beta(k,\chi_2)$ in the relevant wavenumber range $[\kmin,\kmax]$ using some set of simple basis functions, chosen such that the integral in $k$ can be solved analytically. The proposal of \cite{Assassi:2017lea} is to use the FFTlog to perform the power-law decomposition of the relevant functions
\begin{equation}\label{eq_power_law_expansion_explicit}
\mathcal{P}_\mathcal{R}(k) T_\alpha (k,\chi_1) T_\beta (k,\chi_2) = \sumn c_n^{\alpha\beta} (\chi_1,\chi_2) \, k^{\nu_n} \;.
\end{equation}
In this expansion we will refer to $c_n^{\alpha\beta}$ as Fourier coefficients and $\nu_n$ as Fourier frequencies for a given Fourier mode $n$. The details and the practical implementation of this decomposition will be described in the following sections. Let us for the moment assume that such approximation is possible and rewrite the $k$-integral in the following way
\begin{equation} \label{eq_fourier_introduction}
	\begin{split}
		\intinf \frac{dk}{k}  \mathcal{P}_\mathcal{R}(k) \, T_\alpha (k,\chi_1) \bessel{\ell}{k \chi_1} & \, T_\beta (k,\chi_2) \bessel{\ell}{k \chi_2} = \\  \sumn & c_n^{\alpha\beta} (\chi_1,\chi_2) \intinf \frac{dk}{k} \, k^{\nu_n}\,  \bessel{\ell}{k \chi_1} \bessel{\ell}{k \chi_2}\;.
	\end{split}
\end{equation}
If we define
\begin{equation} 
\label{eq_Il}
I_\ell(\nu,t) \equiv 4 \pi \intinf \, \frac{du}{u} \enskip u^{\nu} \, \bessel{\ell}{u}  \bessel{\ell}{u t}\;,
\end{equation}
the angular power spectrum can be rewritten as
\begin{equation}
C_\ell^{\alpha\beta,ij} = \sumn \,\, \intinf d\chi_1 \, W^i(\chi_1) \intinf d\chi_2 W^j(\chi_2) \, c_n^{\alpha\beta}(\chi_1,\chi_2) \, \chi_1^{-\nu_n} \, I_\ell\left(\nu_n , \frac{\chi_2}{\chi_1}\right)~.
\end{equation}
Finally, changing the integration variables to $\chi=\chi_1$ and $t=\chi_2/\chi_1$ we obtain a simplified expression
\begin{equation}
\label{eq_final_cl}
C_\ell^{\alpha\beta,ij} = \sumn  \intinf d\chi \, W^i(\chi)  \intinf dt \, W^j(\chi t) \, c_n^{\alpha\beta}(\chi,\chi t) \, \chi^{1-\nu_n} \, I_\ell\left(\nu_n,t \right)~.
\end{equation}
Notice that this expression is formally the same as in \eqref{eq_old}. However, the spherical Bessel functions are now analytically integrated. The resulting function $I_\ell\left(\nu,t \right)$ has two important properties: First, it has a simple analytical form in terms of hypergeometric functions; secondly, it is smooth, which allows for straightforward numerical integration in $t$. These simplifications are crucial for the efficient numerical evaluation of $C_\ell^{\alpha\beta,ij}$.
\dnew
Another important advantage of decomposition \eqref{eq_power_law_expansion_explicit} is that all cosmology dependence is in the coefficients $c_n^{\alpha\beta}$. This means that the function $I_\ell(\nu,t)$ is cosmology-independent. This provides a motivation to rewrite the expression for $C_\ell^{\alpha\beta,ij}$ such that universal geometrical factors (like the function $I_\ell(\nu,t)$) and cosmology-dependent factors (like the coefficients $c_n(\chi,\chi t)$ and the window functions) are completely separated. To achieve this we can change the order of integrations
\begin{equation}
C_\ell^{\alpha\beta,ij} = \sumn \intinf dt \, I_\ell\left(\nu_n, t \right)  \intinf d\chi \, W^i(\chi)  W^j(\chi t) c_n^{\alpha\beta}(\chi,\chi t) \chi^{1-\nu_n}\;,
\end{equation}
and define a cosmology-dependent function $f_n^{\alpha\beta,ij}(t)$
\begin{equation}
\label{eq_def_fn}
f_n^{\alpha\beta,ij}(t) \equiv \intinf d\chi \, W^i(\chi)  W^j(\chi t) c_n^{\alpha\beta}(\chi,\chi t) \chi^{1-\nu_n}~.
\end{equation}
Using this definition, we can write the angular power spectrum as the following integral
\begin{equation} 
C_\ell^{\alpha\beta,ij} = \sumn \intinf dt \, I_\ell\left(\nu_n ,t \right) f_n^{\alpha\beta,ij}(t)~.
\end{equation}
The main virtue of this expression is separation of geometry and cosmology. The cosmology-dependent function $f_n^{\alpha\beta,ij}(t)$ depends on the observables $\mathcal O_\alpha$ and window functions $W_i$ but it is {\em independent} of the multipole $\ell$. On the other hand, the geometrical function $I_\ell(\nu,t)$ is cosmology-independent and it has to be calculated only once for any cosmology. (In practice it can be stored in a file and loaded when desired.) The function $f_n^{\alpha\beta,ij}(t)$ represent the correlation between the two window functions $W^i$ and $W^j$ weighed by a Fourier mode $n$ represented by $c_n^{\alpha\beta}(\chi,\chi t) \chi^{1-\nu_n}$. The parameter $t$ corresponds to the relative correlation distance in $\chi$. These cosmological Fourier-weighed correlators $f_n^{\alpha\beta,ij}(t)$ are finally convolved with the geometrical Fourier-weighed correlators $I_\ell(\nu_n,t)$ and summed over all Fourier modes $n$ used in the weighing.
\dnew
If we try to compare this more directly to the traditional line-of-sight approach, the integral over $t$ in (\ref{eq_def_fn}) has similarities with the integral over $\chi$
in (\ref{eq_traditional_los}), in which one also convolves a cosmological function  $W^i(\chi)T(k,\chi)$ with a geometrical function $j_\ell(k\chi)$. The final sum over $n$ in (\ref{eq_def_fn}) plays a role similar to the integral over $k$ in (\ref{eq_old}). However, we will see that the new method offers the advantage of better behaved integrals, which results in increased precision and speed.
\dnew 
Before we discuss details and practical implementation of the power-law decomposition and evaluation of $I_\ell(\nu,t)$, let us finish this section by mentioning some symmetry relations which further simplify the expressions above. There are two such relations, which follow simply from a substitution of the integration variables. The geometrical function $I_\ell(\nu,t)$ satisfies
\begin{equation}\label{eq_ilsym}
I_\ell\left( \nu,1/t \right) = t^{\nu} I_\ell\left( \nu, t\right) \;,
\end{equation}
and the cosmological function $f_n^{\alpha\beta,ij}(t)$ obeys
\begin{equation}
f_n^{\alpha\beta,ij}(1/t) = t^{2-\nu_n} f_n^{\beta\alpha,ji}(t)~.
\end{equation}
This allows us to write
\begin{equation} 
	\begin{split}
	C_\ell^{\alpha\beta,ij} & = \sumn \intinf dt \, I_\ell\left(\nu_n ,t\right)  f_n^{\alpha\beta,ij}(t) \\ & = \sumn \left( \int_0^1 dt \, I_\ell\left(\nu_n ,t\right)  f_n^{\alpha\beta,ij}(t) + \int_1^\infty dt \, I_\ell\left(\nu_n ,t\right)  f_n^{\alpha\beta,ij}(t) \right) \\ & = \sumn \left( \int_0^1 dt \, I_\ell\left(\nu_n ,t\right)  f_n^{\alpha\beta,ij}(t) + \int_0^1 du/u^2 \, I_\ell\left(\nu_n ,1/u\right)  f_n^{\alpha\beta,ij}(1/u) \right) \\ & = \sumn \left( \int_0^1 dt \, I_\ell\left(\nu_n ,t\right)  f_n^{\alpha\beta,ij}(t) + \int_0^1 dt \, I_\ell\left(\nu_n , t\right)  f_n^{\beta\alpha,ji}(t) \right)~,
	\end{split}
\end{equation}
leading to the following result
\begin{equation}\label{eq_intxihalf}
C_\ell^{\alpha\beta,ij} = \sumn \int_0^1 dt \, I_\ell\left(\nu_n ,t\right) \left( f_n^{\alpha\beta,ij}(t) +  f_n^{\beta\alpha,ji}(t) \right) ~.
\end{equation}
Alternatively one may use
\begin{equation} \label{eq_intxi}
C_\ell^{\alpha\beta,ij} = \sumn  \int_0^1 dt \, I_\ell(\nu_n ,t) \left(  f_n^{\alpha\beta,ij}(t) +  t^{\nu_n-2}f_n^{\alpha\beta,ij}(1/t) \right)~.
\end{equation}
The difference between the two is how the functions $f_n^{\alpha\beta,ij}(t)$ have to be calculated. Since $C_\ell^{\alpha\beta,ij} = C_\ell^{\beta\alpha,ji}$, their calculation is required only for $i \geq j$. Then in equation (\ref{eq_intxihalf}) one has to calculate the $f_n^{\alpha\beta,ij}$ for all $(i,j)$ combinations, while in equation (\ref{eq_intxi}) only $i\geq j$ are required. 
\dnew 
In practice one has to be careful about evaluating $f_n^{\alpha\beta,ij}(1/t)$ with enough precision. In the integral over $\chi$ in (\ref{eq_def_fn}), any grid in $\chi$ corresponds to a grid in $\chi/t$, which appears as an argument of the functions $W^j$ and $c_n^{\alpha\beta}$. When $t$ becomes very small, one must check that the grid in $\chi/t$ does not become too coarse. Furthermore, we see that $t^{\nu_n-2}$ is divergent for $t \rightarrow 0$ when $\Re[\nu_n]<2$ (which is usually the case, as we shall see in section \ref{sec_powerlaw}). For the case of $f_n^{\alpha\beta,ji}(t)$ these two problems do not arise, and the grid in $\chi t$ is sampled even finer with smaller t. 
\dnew
Both problems can be handled by noting that in many cases $f_n^{\alpha\beta,ij}(1/t)$ only needs to be evaluated for $t$ values far from $0$, such that $1/t$ is never too large. Indeed, for most contributions to the number count spectra like e.g.~density fluctuations or \rsd s, we can assume that the window functions $W^i(\chi)$ have a finite support [${\xmin}^i$,${\xmax}^i$\,]. Then, (\ref{eq_def_fn}) shows that $f_n^{\alpha\beta,ij}(t)$ is non-zero only in the range $[ t_\mathrm{min}^{ij} , t_\mathrm{max}^{ij} ]$ with
\begin{equation}\label{eq_tminij}
t_\mathrm{min}^{ij} = {\xmin}^j/{\xmax}^i~, \qquad t_\mathrm{max}^{ij} ={\xmax}^j/{\xmin}^i ~.
\end{equation}
Thus the function can even vanish entirely in the range $t \in [0,1]$, but only when ${\xmin}^j > {\xmax}^i$. If the redshift bins are arranged in growing redshift and distance order, this usually does not happen for $i \geq j$. However, $f_n^{\alpha\beta,ij}(1/t)$ and $f_n^{\beta\alpha,ji}(t)$ have a support given by
\begin{equation}\label{eq_tminji}
t_\mathrm{min}^{ji} = {\xmin}^i/{\xmax}^j~, \qquad t_\mathrm{max}^{ji} ={\xmax}^i/{\xmin}^j ~,
\end{equation}
and thus vanish entirely for any $t \in [0,1]$ if ${\xmin}^i > {\xmax}^j$, which may happen for some index pairs with $i \geq j$. Even if neither function vanishes entirely, the minimum of $t_\mathrm{min}^{ij}$ and $t_\mathrm{min}^{ji}$ provides an overall lower limit on $t$. We will also see in section (\ref{sec_bessel}) that the functions $I_\ell(\nu,t)$ will provide additional limits on \tmin. 
The existence of such a minimum value of $t$, and thus of a maximum value of $1/t$, tells us that we can perform integrals in the form of (\ref{eq_intxi})  without worrying too much about sampling or divergence issues in the limit $t\to 0$.
\dnew
For other types of contributions to the spectra like e.g. weak lensing, we will see that the window functions are defined on a support starting from $\chi=0$. In that case the previous discussion does not apply and we will go back to the integral in the form of (\ref{eq_intxihalf}).
\dnew
In the practical implementation of the new method, we actually calculate the function $f_n^{\alpha\beta,ij}(t)$ using a rather coarse grid in $t$ since it is very smooth. Only when convolving with the more oscillating function $I_\ell(\nu,t)$ will a more fine-grained grid be required. We will then interpolate within the coarse grid of $f_n^{\alpha\beta,ij}(t)$ values using Cubic Hermite Spline Interpolation.

\subsection{The Power-law (FFTlog) Decomposition}\label{sec_powerlaw}
In this section we will give details of the power-law decomposition in \eqref{eq_power_law_expansion_explicit} and the FFTlog algorithm used to achieve it. The power-law expansion of a function $f(\vx,k)$ is simply a Fourier decomposition in $\log(k)$. To see this let us consider a generic function $f(\vx; k)$, periodic in the interval $[\kmin,\kmax]$ with the period $T\equiv \log(\kmax)-\log(\kmin) = \log(\kmax/\kmin)$. The vector $\vx$ describes an arbitrary set of extra variables. The coefficients of the logarithmic Fourier expansion of the function $f(\vx,k)$ are given by
\begin{equation}
c_n(\vx) = \frac{1}{T} \int\limits_{\log(\kmin)}^{\log(\kmax)} f(\vx,k) \exp\left(- \frac{2 \pi i \, n}{T} \log(k) \right) d\log(k)~.
\end{equation}
Using a finite set of these Fourier coefficients we can approximate the original function as 
\begin{equation} \label{eq_power_law_expansion0}
\bar f(\vx; k) = \sumn c_n(\vx) \exp\left[\frac{2 \pi i\, n}{\log(\kmax/\kmin)} \log(k) \right] = \sumn c_n(\vx) \, k^{\nu_n}~,
\end{equation}
which can be easily rewritten in the desired form
\begin{equation} \label{eq_power_law_expansion}
\bar f(\vx; k) = \sumn c_n(\vx) \, k^{\nu_n}~, \qquad {\rm where} \qquad \nu_n = \frac{2\pi\, i \, n}{\log(\kmax/\kmin)} \;.
\end{equation}
The more terms are kept in this sum, the closer $\bar f(\vx,k)$ is to the function $f(\vx,k)$. In practice, the Fourier coefficients $c_n(\vx)$ can be computed binning the function in $\log(k)$ and using the Fast Fourier Transform (FFT) algorithm --- hence the name \tquote{FFTlog} commonly used in cosmology literature \cite{Hamilton:1999uv,Assassi:2017lea,Gebhardt:2017chz,Schmittfull:2016jsw,McEwen:2016fjn} (see Appendix~\ref{ap_powerlaw} for further details on the relation to the FFT). 
\dnew
One immediate question is why this method is applicable to the case we are interested in, where the transfer functions or the primordial power spectrum are not periodic functions in $k$. The reason is that these smooth functions in $k$ are always multiplied by the spherical Bessel functions which peak for $k\sim \ell/\chi$ and decay both in $k\to 0$ and $k\to \infty$ limit. Therefore, the integral picks up most of its contribution from a finite range of scales and replacing the true function with its approximation \eqref{eq_power_law_expansion0} leads to negligible error even when the limits of integration are taken to be $0$ and $\infty$. However, the asymptotic behavior of the spherical Bessel functions in two different limits is such that they do not approach zero equally fast
\begin{equation}
j_\ell(k\chi) \to k^\ell\;, \quad k\to0\;, \qquad {\rm and} \qquad j_\ell(k\chi) \to k^{-1}\;, \quad k\to\infty\;.
\end{equation}
Depending on the observable and the form of the transfer functions, the result may be much more sensitive to high or low $k$. Therefore, to ensure better convergence properties, it is convenient in practice to use a slightly more general form of the Fourier transform which allows additional freedom in regulating the asymptotic behavior. This can be simply achieved by writing $k^b\cdot k^{-b}f(\vx,k)$ and applying the logarithmic Fourier expansion to $k^{-b}f(\vx,k)$ instead of $f(\vx,k)$. In this way, one can always choose the ``tilt'' $b$ to ensure more symmetric behavior of the integrand in the two limits. Effectively, this procedure shifts the Fourier frequencies $\nu_n$ in the following way
\begin{equation} \label{eq_power_law_expansion_nu}
\enskip \nu_n = \frac{2\pi\, i \, n}{\log(\kmax/\kmin)}  +b~.
\end{equation}
We will give more details on the choice of $b$ for different observables below. 
\dnew
The best convergence in \eqref{eq_power_law_expansion} is reached when the function that we want to Fourier transform (i.e. $k^{-b}f(\vx,k)$) in the range $[\kmin, \kmax]$ has equal values at the boundaries,
\begin{equation}\label{eq_continuity}
\kmin^{-b} f(\vx,\kmin) = \kmax^{-b} f(\vx,\kmax) ~,
\end{equation}
because it can then be considered as a sample of a periodic and continuous function in the range $[\kmin, \kmax]$. Our strategy is choosing $b$ is such that $k^{-b} f(k) \longrightarrow 0$ on both ends of the interval. For the values of $\kmin$ and $\kmax$\,, we should in principle consider the finite range of scales to which  the $C_\ell$'s that we want to compute are actually sensitive through their convolution kernels. Then $\kmin$ should be of the order of the scale crossing the Hubble radius today, $k_0 \sim a_0 H_0$, and $\kmax$ should depend on the maximum multipole and minimum redshift considered. Typically, one is interested in multipoles such that the non-linear corrections are either negligible, or small enough to be well under control: this typically limits the sensitivity of the $C_\ell$'s to maximum wavenumbers of the order of 1 to 10 inverse Megaparsecs.  We will come back to the details of choice of $\kmin$ and $\kmax$ in later sections. 
\dnew
Given the previous discussion, we must ensure that $k^{-b} f(\vx,k)$ has vanishing values at the boundaries of the interval $[\kmin, \kmax]$, where $f(\vx,k)$ is the left-hand side of equation (\ref{eq_power_law_expansion_explicit}): it is simply the product of the dimensionless primordial spectrum ${\cal P}_{\cal R}(k)$ and two transfer functions $T_\alpha(k,\chi)$. The behaviour of this function in the small-scale and large-scale limits is known at least approximately. Observations confirm that the dimensionless primordial spectrum is equal or close to a power-law of the form $k^{n_s-1}$\,. The transfer functions depend on the gauge and on the type of observable being computed. Here we discuss two different relevant cases.
\dnew
\textit{Density transfer functions.} In the important case of density fluctuations, which usually dominate the number count spectra, $T_\alpha(k,\chi)$ has a well-known behaviour as a function of $k$. For any time after radiation-matter equality, it grows like $k^2$ between $k_0$ and the scale that crossed the Hubble radius at Matter-Radiation equality, $k_\mathrm{eq}\sim 10^{-2}h/$Mpc. At the level of linear theory, it then keeps growing logarithmically on scales $k>k_\mathrm{eq}$. Non-linear corrections slightly increase the slope beyond the scale of non-linearity. Strong non-linear effects like baryonic feedback occur on even larger wavenumbers and are irrelevant for the range of multipoles usually used for cosmological parameter inference. The behavior that we just described is valid in any gauge, since gauge differences only show up on super-Hubble scales which are outside of our interval. Note that this precise behavior of the transfer function in the range $k_\mathrm{eq}<k<\kmax$ is not essential in our discussion, since our purpose is just to choose the tilt $b$ such that $k^{-b}f(\vx,k)$ is suppressed in $\kmax$ compared to its peak value. 
\dnew
In summary, the asymptotic behavior of the function $k^{-b}f(\vx,k)$ is typically given by
\begin{equation}
	\begin{aligned}
		k^{-b}f(\vx,k) &\sim k^{n_s+3-b}    & (k \ll k_\mathrm{eq})~,\\
		k^{-b}f(\vx,k) &\sim k^{n_s-1-b} \ln(k)^2 \qquad  &(k \gg k_\mathrm{eq})~.
	\end{aligned}
\end{equation}
Thus any tilt factor in the range $n_s-1 < b < n_s+3$ will ensure that (\ref{eq_continuity}) is fullfiled, with vanishing values at the boundaries compared to the peak value.
\dnew
The choice of tilt is also important for the calculation of the geometrical function $I_\ell(\nu_n,t)$ given by (\ref{eq_Il}). Indeed, the tilt $b$ enters the definition of the coefficients $\nu_n$ in equation (\ref{eq_power_law_expansion_nu}). 
Thus it is important to choose $b$ such that within this range, the $u$ integral of equation (\ref{eq_Il}) is well behaved, such that $I_\ell(\nu_n,t)$ can be calculated. This condition was already investigated in \cite{Assassi:2017lea} and imposes a range $-2 \ell < \Re[\nu_n] < 2$. For $\Re[\nu_n]=b$, this overlaps with the previous range $n_s-1 < b < n_s+3$, for whatever realistic value of $n_s$. The optimal choice is found to lay somewhere between $1.5$ and $1.9$.
\dnew
However, with $n_s \simeq 1$, $\kmax \simeq 1 \,\, h/\text{Mpc}$ and $b \simeq 1.9$\,,\, the value of $k^{-b} f(\vx, k)$ at $k_\mathrm{max}$ is not very small: it is around $10\%$ of the peak value. Stopping the integral around this scale would imply that we FFT-transform a periodic function with a discontinuity in $k_\mathrm{max}$. The FFT method is very sensitive to such discontinuities. If nothing is done to further reduce the difference between the values at the edge, 
this small discontinuity generates a small unphysical contribution to the FFT coefficient $c_n$ and leads to a percent-level error in the final result. To solve this problem, one can introduce additional regulatory mechanisms. Knowing that for the relevant $C_\ell$'s, the precise form of the transfer function is not important for wavenumbers bigger than about one inverse Megaparsec, one has a lot of freedom for regularizing the function $k^{-b}f(\vx, k)$ in the large $k$ limit.
The choice of reference \cite{Assassi:2017lea} was to numerically evolve wavenumbers up to a $\kmax$ of $\sim 52h/\text{Mpc}$, and to multiply the transfer functions by an exponential cut-off around $k_\mathrm{cut} \sim 10h/\text{Mpc}$. In the \texttt{CLASS} implementation, we choose a computationally faster method: we extend the upper boundary in the integrals 
to about $\kmax \simeq 10^3h/\text{Mpc}$, but with an analytic extrapolation of the transfer function of the form $T(k) \sim \log(a k)$ above the maximum wavenumber for the numerical evaluation, set by default by \texttt{CLASS} for a redhsift bin of mean conformal distance $\chi_i$ to $k_\mathrm{num} = 2.4 \, \ell_\mathrm{max} / \chi_i$ (typically  of the order of $1 h/\text{Mpc}$ ). We stress that this extrapolation is not meant to be physical and is just a regulatory artifact. We tested different values of $\kmax$ in the range from $10h/\text{Mpc}$ to $10^3h/\text{Mpc}$ and found that the result was very stable, varying by less than one permille.
\dnew
\textit{Other transfer functions.} We will present in section~\ref{sec_behaviour_k} a list of transfer functions which are useful for the calculation of number count spectra (with all GR corrections) and weak lensing spectra. We will see that the other transfer functions scale with $k$ roughly in the same way as the density transfer function, but with an extra gobal factor $k^{-2}$. In that case, our strategy consists in doing an FFTlog transformation of \mbox{$\bar{T}_\alpha(k,\chi)=k^2 T_\alpha(k,\chi)$}, which then has the same behavior in $k$ as the density contribution. For this, we write \mbox{$T_\alpha(k,\chi)=k^{-2} \bar{T}_\alpha(k,\chi)$}, and denote the FFT transform of the product $\mathcal{P}_\mathcal{R}(k) \bar{T}_\alpha(k,\chi_1) \bar{T}_\beta(k,\chi_2)$ as $\bar{c}_n^{\,\alpha\beta} (\chi_1,\chi_2)$. This changes equation (\ref{eq_fourier_introduction}) to
\begin{equation}
	\begin{split}
		\int_0^\infty \frac{dk}{k} \, k^{-4}\, \mathcal{P}_\mathcal{R}(k) \, \bar{T}_\alpha(k,\chi_1) \bessel{\ell}{k \chi_1}\, & \bar{T}_\beta(k,\chi_2) \bessel{\ell}{k \chi_2}  = \\  \sumn & \bar{c}_n^{\,\alpha_\beta}(\chi_1,\chi_2) \int_0^\infty \frac{dk}{k} \, k^{\nu_n-4}\,  \bessel{\ell}{k \chi_1} \bessel{\ell}{k \chi_2}~.
	\end{split}
\end{equation}
Thus, this redefinition effectively replaces $\nu_n$ with $\nu_n-4$ or $\nu_n-2$, both in the definition of $f^{ij}(t)$ in equation (\ref{eq_def_fn}) and in the argument of $I_\ell(\nu,t)$ in equations (\ref{eq_intxihalf}, \ref{eq_intxi}). This implies that the values of the tilt such that the integrals $I_\ell(\nu,t)$ are well behaved now shift to either $-2 \ell +2 < b < 4$ or $-2 \ell +4 < b < 6$. In all these cases and for $\ell \geq 2$, the tilt value chosen in the density case, $b \simeq 1.9$, is still valid, and the Bessel integrals are well behaved.
\dnew
Finally, let us note that the reference \cite{Assassi:2017lea} proposes an additional trick to shift the tilt $b$ by using an integrations by part. In the present implementation in \texttt{CLASS} we did not experience the need to use this technique. It is further discussed in section \ref{sec_behaviour_k}.

\subsection{The Analytical Bessel Integrals} \label{sec_bessel}
We conclude the discussion on the method by making a few comments on evaluation of Bessel integrals $I_\ell(\nu,t)$. While \cite{Assassi:2017lea} and \cite{Gebhardt:2017chz} provide explanations of how to obtain $I_\ell(\nu,t)$ in a fast and accurate manner, we focus here on several aspects raised when working with finite precision arithmetics.
\dnew
The $I_\ell(\nu,t)$'s can be expressed in terms of Hypergeometric functions for $t<1$,
\begin{equation} \label{eq_bessel_int_hyper2f1}
I_\ell(\nu,t) = \frac{2^{\nu-1} \pi^2 \Gamma(\ell+\frac{\nu}{2})}{\Gamma(\frac{3-\nu}{2})\Gamma(\ell+\frac{3}{2})} \;\; t^\ell \;\;\hyp{\frac{\nu-1}{2}}{\ell+\frac{\nu}{2}}{\ell+\frac{3}{2}}{t^2}~.
\end{equation}
For $t>1$ they can be calculated using equation (\ref{eq_ilsym}). From this equation it follows that $I_\ell(\nu,t)$ vanishes for $t \rightarrow 0$.  For $t=1$ the hypergeometric function can be expressed in terms of gamma functions. The typical behaviour of $I_\ell(\nu,t)$ is such that its absolute value peaks at $t=1$. Furthermore, given that $I_\ell(\nu,t)\sim t^\ell$, for high values of $\ell$ the Bessel integrals have support only in a narrow interval close to $t=1$. This means that in practice, in order to reach a given precision, we can find some $\tmin$ such that $|I_\ell(\nu,\tmin)|<\epsilon |I_\ell(\nu,1)|$,  where $\epsilon \ll 1$, and approximate the integrals (\ref{eq_intxihalf}, \ref{eq_intxi}) as ranging from $\tmin$ to 1. In this way $I_\ell(\nu,t)$ does not need to be evaluated for $t<\tmin$, and the integrals over $t$ only need to be carried over a small range, which gets narrower with larger $\ell$'s.
\dnew 
For very large $\ell$ the result obtained using the Bessel integrals must agree with the Limber approximation \cite{1953ApJ...117..134L,Kaiser:1996tp,Lemos:2017arq}, 
\begin{equation}
\bessel{\ell}{u}  \bessel{\ell}{u t} \approx \frac{2\pi^2}{\ell_0} \delta(\ell_0-u) \delta(\ell_0-t u)\;, \qquad \mathrm{with} \qquad \ell_0=\ell+1/2~.
\end{equation}
This would imply
\begin{equation} \label{eq_limber}
I_\ell(\nu,t) \approx 2\pi^2 \,({\ell_0\,})^{\nu-3} \,\delta(1-t)~,
\end{equation}
which is indeed consistent with the result of the explicit computation of $I_\ell(\nu,t)$ in this limit, using the fact that $(1-\tmin)$ is proportional to $\ell^{-1}$ and $I_\ell(\nu,1)$ proportional to $\ell^{\nu-2}$ (see appendix \ref{ap_traffo}). However, our method is such that the Limber approximation never has to be used. Once $f_n^{ij}(t)$ has been evaluated in the range $0<t<1$ for the purpose of computing the low multipoles with the full integral of (\ref{eq_intxihalf}, \ref{eq_intxi}), the computational cost of the integrals for higher multipoles is minor, because the support of integration $[t_\mathrm{min},1]$ gets smaller. Thus there would be not much gain in using the Limber approximation and in our \texttt{CLASS} implementation we currently choose not to use it for any $\ell$. 
\dnew
The direct evaluation of the Bessel integrals on the whole interval $0<t<1$ using equation \eqref{eq_bessel_int_hyper2f1} or similar other representations (see appendix \ref{ap_traffo}) can be relatively slow and inaccurate. The reason is that the power series of the Gauss hypergeometric function ${}_2F_1$ does not converge well in finite precision arithmetics due to large cancellations and/or many terms in the series have to be kept. This problem is particularly relevant for large imaginary parameters. Therefore, it is important to avoid direct evaluation whenever possible. Luckily, there are several alternative methods to efficiently calculate the Bessel integrals which we list below. On the basis of several speed and accuracy tests (performed by comparing with reference results provided by the python library \texttt{mpmath}), we have established and implemented in our code a list of threshold values $t_i$ in the range $[0,1]$ at which it is advantageous to switch from one method to another. These thresholds depend both on $\nu$ and $\ell_\mathrm{max}$\,.
\dnew
\textit{Taylor method.} For $t\rightarrow0$, we can use an analytic Taylor expansion of the hypergeometric function in the vicinity of zero. A similar approach can be taken for $t\rightarrow1$, because the hypergeometric function can first be transformed with a change of variable $t^2 \rightarrow \left((1-t^2)/(1+t^2)\right)^2$, and then expanded in small $\epsilon=1-t$. This transformation only works for tiny $\epsilon$ values, otherwise the result depends on precise cancellations between two hypergeometric functions. The different transformations for the $I_\ell(\nu,t)$'s are listed in appendix \ref{ap_traffo}. The Taylor method is not as fast as the recursion methods discussed next, but it is more robust in the close vicinity of the edges. We use it in two small ranges $0 < t\leq t_1$ and $t_2 < t < 1$.
\dnew
\textit{Recursion methods.} The Bessel integral $I_\ell(\nu,t)$ obeys the recursion relation \cite{Assassi:2017lea}
\begin{equation}
\left(3+\ell-\frac{\nu}{2}\right) I_{\ell+2}(\nu,t) = \frac{1+t^2}{t} \left(\ell+\frac{3}{2}\right) I_{\ell+1}(\nu,t) - \left(\ell+\frac{\nu}{2}\right) I_\ell(\nu,t)~,
\end{equation}
where the $I_0(\nu,t)$ and $I_1(\nu,t)$ are known analytically: 
\begin{align}
	I_0(\nu,t) &= 2\pi \cos\left(\frac{\pi \nu}{2}\right) \Gamma(\nu-2) t^{-1} \left[(1+t)^{2-\nu}-(1-t)^{2-\nu}\right]~, \\
	I_1(\nu,t) &= 2\pi \frac{ \cos\left(\frac{\pi \nu}{2}\right)\Gamma(\nu-2) }{(4-\nu) t^2} \left[(1+t)^{2-\nu}((1+t)^2+\nu t)-(1-t)^{2-\nu}((1+t)^2-\nu t)\right]~. \nonumber
\end{align}
We can follow the recursion relation in forward direction $\ell \rightarrow \ell+1$, starting from $\ell=0$ and $\ell=1$. Alternatively the recursion relation can be used in the backward direction $\ell \rightarrow \ell-1$, starting from some $\ell_{seed}$ and $\ell_{seed}-1$. The forward and backward recursions are always the fastest methods but they are not always stable.
\dnew 
The method is only stable until a maximum value of $\ell$ which increases with $t$. We employ it only above a threshold value $t_3$ such that stability extends up to at least to the highest multipole value $\ell_\mathrm{max}$ needed for the angular spectrum computation.
\dnew
For $t<t_3$\,, we try to use the backward recursion whenever possible, starting from some $I_{\ell,seed}(\nu,t)$ and $I_{{\ell,seed}-1}(\nu,t)$ given by the direct calculation (\ref{ap_eq_hight}). We mentioned that this calculation can be slow and inaccurate, but high accuracy is not needed in this context. Indeed, the error made on the initial terms is reduced at each iteration. Therefore, instead of starting from the highest multipole value $\ell_\mathrm{max}$ needed for the angular spectrum computation, we start from a value offset by some amount $\Delta \ell$\,;\, $\ell_{seed}=\ell_\mathrm{max}+\Delta\ell$, based on Miller's Recurrence Algorithm. The error on $I_{\ell,max}(\nu,t)$ is then reduced by a factor $t^{\Delta\ell}$. The optimal amount of offset $\Delta\ell$ depends strongly on the precision with which the initial values have been computed, but also on $t$ and $\nu$ (for instance, we found that it should scale as $(1-t)^{-1}$). We implemented in the code an ansatz for $\Delta\ell(\nu, t)$, keeping in mind that it is usually much faster to take a larger offset than to calculate the initial seeds with higher precision. Furthermore, we can use a trick similar to the one presented in \cite{Gebhardt:2017chz}: Once we reach the analytically known $\ell=0$, we can compare the result obtained by backward recursion with the analytic result, calculate the complex ratio
\begin{equation}
\lambda(\nu,t) = \frac{I_0^{(\mathrm{analytical})}(\nu,t)}{I_0^{(\mathrm{recursion})}(\nu,t)}~,
\end{equation}
and multiply a posteriori all $I_\ell(\nu,t)$ with this ratio.  
\dnew
The backward recursion is more stable for smaller $t$ values (but not very close to zero, which is why we use the Taylor method below $t_1$). We find that for some values of $\nu$, the backward recursion gives good results in the whole range from Taylor to Forward recursion, $t_1 < t < t_3$, allowing us to switch directly from the backward to the forward method at $t_3$\,. For some other values of $\nu$, the backward recursion is only accurate up to another threshold value $t_2$, and the last remaining range from Backward to Forward recursion, $t_2 < t < t_3$\,, needs to be covered by a last method.
\dnew
\textit{Helper function method.} Finally, in regions $t_2 < t < t_3$ where neither the recursion relations nor the Taylor expansions give a satisfyingly small error, we switch to the method introduced in Appendix E of \cite{Gebhardt:2017chz}, which introduces what we call a \mbox{\tquote{helper function}} for the purpose of computing the hypergeometric function. We have adapted their approach for our purposes. For instance, we do not consider different $\ell\neq\ell'$ in the Bessel functions, we use a different method for setting the seeds, and we use a different forward-backward recursion switching criteria. This method is not as fast as the recursion methods, but it appears to be almost universally convergent.
\dnew
For the Taylor and \tquote{helper function} methods, it is important to exploit the recursion relations for the Gamma functions $\Gamma(x+1)=x\Gamma(x)$ as much as possible. Otherwise, the evaluation of a few $\Gamma(x)$ from scratch in each recursion step would be prohibitively slow.
	\section{Different Source Terms}\label{sec_types}

Reference \cite{Assassi:2017lea} already addressed the fact that the transfer function of the number count spectrum involves multiple terms, and focused in particular on {\rsd}s. Here we want to go into the full details of all the terms and problems involved. The derivation of all the necessary source terms for the galaxy number count is given in \cite{Bonvin:2011bg} and their concrete expressions in \texttt{CLASS} are summarized in \cite{DiDio:2013bqa} (on page 19). We also give a summary of these terms (with additional information relevant to the present work) in Appendix \ref{ap_types}, tables \ref{tab_terms} and \ref{tab_gauges}.
\dnew 
We write the total transfer function of number count in the form
\begin{equation}
	T(k,\chi) = \sumx T_x(k,\chi)~,
\end{equation}
where $x$ labels the different contributions: Doppler terms and {\rsd}s, lensing terms, and other gravitational terms (accounting for small GR corrections).\footnote{Note that we have used labels $\alpha$ and $\beta$ on the transfer function in the previous sections to indicate different kind of observables in the general cross-spectra. The latin indices $x$ and $y$ here indicate different contributions to a single observable. For clarity we will suppress Greek indices in this section, but it is always assumed that our formulas apply to a generic cross spectrum.} 
Consequently we have to introduce a double sum in the definition of the function $f_n^{ij}(t)$,
\begin{equation}
	f_n^{ij}(t) = \intinf d\chi \,  W^i(\chi)W^j(\chi t)  \sumxy c_n^{xy}(\chi,\chi t) \;,
	\label{eq_fijxy}
\end{equation}
with $c_n^{xy}(\chi,\chi t)$ defined analogously to equation (\ref{eq_power_law_expansion_explicit}), using
\begin{equation}
	\mathcal{P}_\mathcal{R}(k) T_x(k,\chi_1) T_y(k,\chi_2) = \sumn c_n^{xy}(\chi_1,\chi_2) \, k^{\nu_n}~.
\end{equation}
Since the $C_\ell$'s depend linearly on the Fourier coefficients $c_n^{xy}$, they can also be decomposed as $C_\ell^{ij} = \sumxy C_\ell^{ij,xy}$.
\dnew
While most terms can be cast in the general form of equations (\ref{eq_fijxy}) and (\ref{eq_intxihalf}, \ref{eq_intxi}) easily, there are two kinds of source terms for which this is possible, but not straightforward. These are terms for which the transfer function $\Delta_{\ell}^i(k)$ would involve not a spherical Bessel function, but its first or second derivative (Doppler and {\rsd} terms); and those for which it contains one additional integral over time or comoving radius (weak lensing and some other GR corrections). In what follows we give details of how to deal with each of these complications separately, since they involve different challenges and solutions.

\subsection{Derivatives of Bessel Functions}\label{sec_derivs}

The Doppler terms and {\rsd}s include derivatives of the Bessel functions. They give rise to contributions of the form 
{\small\vspace*{0.5em}
\begin{equation}
C_\ell^{ij,xy} = 4 \pi \sumn \intinf d\chi_1 d\chi_2 \, W^i(\chi_1) W^j(\chi_2) \, c_n^{xy}(\chi_1,\chi_2) \,  
 \intinf \frac{dk}{k} \, k^{\nu_n}\,  j_{\ell}^{(n_1)}(k \chi_1) j_{\ell}^{(n_2)}(k \chi_2)~,
\end{equation}
}
where $n_1, n_2 = 0,1,2$ and $j_\ell^{(0)}$, $j_\ell^{(1)}$, $j_\ell^{(2)}$ denote respectively the zero-, first- and second-order derivatives of the spherical Bessel function. We will make use of integrations by parts to obtain an expression without derivatives of the Bessel functions.\footnote{Note that, in principle, we could avoid any integration by parts and instead use the relation between derivatives of Bessel functions and linear combinations of $\bessel{\ell}{x}$ and $\bessel{\ell+1}{x}$. This would require the introduction of additional analytical integrals, which do not obey the same transformation properties that make $I_\ell(\nu,t)$ relatively easy to calculate (see appendix \ref{ap_traffo} for a more thorough discussion). Thus, the integration by parts method turns out to be a simpler approach.} The boundary terms in the integration by parts vanish since
\begin{equation} \label{eq_boundary}
	\lim\limits_{x\to 0} \bessel{\ell}{x} = \lim\limits_{x\to \infty} \bessel{\ell}{x}= 0 \qquad \text{for} \quad l\geq 1~.
\end{equation}
In the traditional line-of-sight approach, one could write that 
\begin{equation}
\Delta_\ell^{i,x}(k) =	\intinf d\chi W^i(\chi) T_x(k,\chi) \frac{\partial\bessel{\ell}{k \chi}}{\partial k\chi} = - \intinf d\chi \frac{\partial \left(W^i(\chi) T_x(k,\chi) \right)}{k\, \partial \chi} \bessel{\ell}{k \chi} ~.
\end{equation}
We see that each derivative of a Bessel function both gives us a power of $k$ and a time derivative of the product of the window and transfer function. We can use exactly the same approach in the FFT formalism and, for instance, in the case $(n_1, n_2)=(1,0)$ write
{\small
\begin{equation}
\label{eq_with_Derivs}
C_\ell^{ij,xy} = 4 \pi \sumn \intinf d\chi_1 d\chi_2 W^j(\chi_2) \, \frac{\partial\left(W^i(\chi_1) c_n^{xy}(\chi_1,\chi_2)\right)}{\partial \chi_1}  
\intinf \frac{dk}{k} \, k^{\nu_n-1}\,  j_{\ell}(k \chi_1) j_{\ell}(k \chi_2)~.
\end{equation}
}
\dnew
While taking derivatives of the Fourier coefficients $c_n^{xy}(\chi_1,\chi_2)$ is possible, it would be very slow in practice, because these derivatives should be evaluated in each point of the discrete $(\chi_1, \chi_2)$ grid. This problem is amplified by the fact that we also need all other derivatives of the form $(\partial_{\chi_1})^{n_1} (\partial_{\chi_2})^{n_2} c_n^{xy}(\chi_1,\chi_2)$. These issues could be avoided if the coefficients $c_n^{xy}$ were separable in $\chi_1$ and $\chi_2$, and of the form
\begin{equation}\label{eq_separability}
	c_n^{xy}(\chi_1,\chi_2) = D^x(\chi_1)D^y(\chi_2) c_n ~,
\end{equation}
because then the derivatives could be calculated independently. This \tquote{separability} ansatz was implicitly suggested in \cite{Assassi:2017lea,Gebhardt:2017chz}. It relies on the separability of the transfer function into time and wavenumber dependence, $T_x(k,\chi) = D_x(\chi) T_x(k,0)$. This assumption is only fulfilled in a \lcdm\, universe without massive neutrinos, in the sub-Hubble limit and within linear theory. However, the deviation from the \tquote{separable} limit remains small even in the presence of massive neutrinos, non-linear corrections or other physical ingredients leading to a scale-dependent growth factor. We can take advantage of this and define rescaled Fourier coefficients
\begin{equation}\label{eq_semi_separability}
\tilde{c}_n^{xy}(\chi_1,\chi_2) = \frac{c_n^{xy}(\chi_1,\chi_2)}{ D_x(\chi_1)D_y(\chi_2)}  ~,
\end{equation}
which are only weakly dependent on conformal time. We obtain the effective growth factor $D_x(\chi)$ by extracting it from the exact transfer function $T_x(k,\chi)$. To focus on the scales of interest, we weigh the contributions from different positions in the $k$-grid differently using weights $w_m$,
\begin{equation}
D_x(\chi) = \frac{\sum_m w_m T_x(k_m,\chi)/T_x(k_m,0)}{\sum_m w_m}~.
\end{equation}
For the weights $w_m$ we choose a Gaussian in $\log(k)$, centered around some scale of interest $k_0$\, ($=1 h/\text{Mpc}$ in our implementation), with a standard deviation of half-a-decade. The precise form of the weights and the scale $k_0$ do not influence our results beyond a relative deviation of $10^{-6}$.
\dnew
The rescaled Fourier coefficients of equation (\ref{eq_semi_separability}) allow us to work in a \tquote{semi-separable} approximation in which we assume that the dependence of the factors $\tilde{c}_n^{xy}(\chi_1,\chi_2)$ on conformal time is much weaker than that of the growth factors $D^x(\chi)$. In this limit the derivative is now \textit{approximated} by
\begin{equation}
\begin{split}
\frac{\partial}{\partial \chi_1}c_n^{xy}(\chi_1,\chi_2) & = \frac{\partial}{\partial \chi_1} \Big( D_x(\chi_1)D_y(\chi_2) \tilde{c}_n^{xy}(\chi_1,\chi_2) \Big) \\ & \approx D_y(\chi_2) \frac{\partial D_x(\chi_1)}{d\chi_1} \tilde{c}_n^{xy}(\chi_1,\chi_2)~,
\label{eq_semi_separability_derivative}
\end{split}
\end{equation}
effectively giving us the same factorization as when assuming full separability. This allows us to write the terms with derivatives of the spherical Bessel functions in a simple form. For instance, in the example in equation (\ref{eq_with_Derivs}), the expression for the angular power spectrum can be written in the form
\begin{equation}\label{eq_WD}
\begin{split}
C_\ell^{ij,xy}  = \sumn \intinf d\chi_1 d\chi_2  \, &\frac{\partial\left(W^i(\chi_1) \, D_{x}(\chi_1)\right)}{\partial \chi_1} \,  W^j(\chi_2)  D_{y}(\chi_2) \\ & \qquad \qquad \times \, \tilde{c}_n^{xy}(\chi_1,\chi_2) \,
\chi_1^{-(\nu_n-1)} I_\ell\left(\nu_n-1, \frac{\chi_2}{\chi_1}\right)~.
\end{split}
\end{equation}
This can be evaluated very efficiently, since the derivatives of the products $W^i D_x$ can be pre-calculated and later interpolated at some $\chi$ and $\chi t$ (and additionally $\chi/t$ for equation (\ref{eq_intxi})\,). 
This approach applies to any combination of derivatives.
The precision tests presented in section~\ref{sec_concl} prove that the semi-separable approximation of equation~(\ref{eq_semi_separability_derivative}) does not introduce any sizable inaccuracy, even when non-linear corrections are implemented (with fitting formulas like Halofit \cite{Smith:2002dz,Takahashi:2012em}), or when massive neutrinos are introduced (see section \ref{sec_performance}).

\subsection{Integrals over Bessel Functions}

Some source terms accounting for weak lensing and other GR corrections involve an additional integral over conformal time, such that in the traditional line-of-sight approach their harmonic transfer function reads
\begin{equation}
\Delta_\ell^{i,x}(k) =	\int_0^{\tau_0} d\chi' \, W^i(\chi') \, \int_0^{\chi'} d \chi \, f(\chi,\chi') \, T_x(k,\chi) \, \bessel{\ell}{k \chi}~,
\end{equation}
where $\tau_0$ is the conformal age of the universe, and $f(\chi,\chi')$ is a convolution kernel. In the case of lensing, $\chi'$ would be the distance to the source and $\chi$ the distance to the lens. To obtain the same functional form as other harmonic transfer functions we can swap the order of integration \cite{DiDio:2013bqa},
\begin{equation}
\Delta_\ell^{i,x}(k) =	\int_0^{\tau_0} d \chi \int_{\chi}^{\tau_0} d\chi' \, W^i(\chi') \, f(\chi,\chi') T_x(k,\chi) \, \bessel{\ell}{k \chi}\,~,
\end{equation}
leading to the definition of an new window function
\begin{equation}
\label{eq_lens_window}
\tilde{W}(\chi) = \int_{\chi}^{\tau_0} d\chi' \, W^i(\chi') \, f(\chi,\chi')~.
\end{equation}
Once again, the same approach can be readily transposed to the FFT method. For instance, the cross-correlation between one integrated and one non-integrated term reads
\begin{equation}
C_\ell^{ij,xy} = \sumn 
\int_0^{\tau_0} d\chi_1 \, \tilde{W}^i(\chi_1) \int_0^{\tau_0}  d\chi_2 \, W^j(\chi_2) \, c_n^{xy}(\chi_1,\chi_2) \,  
\chi_1^{-\nu_n} \, I_\ell \hspace*{-2.5pt}\left(\nu_n, \frac{\chi_2}{\chi_1}\right)~,
\end{equation}
with $\tilde{W}^i(\chi)$ defined like in equation (\ref{eq_lens_window}).
\dnew
The important difference compared to non-integrated terms comes from the broad support of the integrated window function $\tilde{W}^i(\chi)$\,. If we assume $W^i(\chi)$ is non-zero in the range $[\chi^i_\mathrm{min}, \chi^i_\mathrm{max}]$, then $\tilde{W}^i$ has support in the whole $[0, \chi^i_\mathrm{max}]$ range.
Thus $\tilde{W}^i(\chi)$ has to be sampled with more points in $\chi$, an effect that is worsened by the oscillatory nature of the $\chi^{1-\nu_n}$ factor appearing in the definition of $f_n^{ij}(t)$ in equation (\ref{eq_def_fn}).
\dnew
For contributions involving one integrated and one non-integrated term, the definition of $\chi$ becomes important.
It is better to define $\chi$ as the conformal distance of the non-integrated term.
In that case $\chi$ still has a restricted support, allowing for evaluation with a sparser grid. 
The range of $t$ also has to be adjusted; the new range is $[0,1]$, as can be seen from equations (\ref{eq_tminij}) and (\ref{eq_tminji}). 
\dnew
For contributions involving two integrated terms, $\chi$ has a broad support in any case. Not only are more sampling points required, but since small $\chi$ values are allowed, the oscillatory nature of the $\chi^{1-\nu_n}$ factor in the definition of $f_n^{ij}$ in equation (\ref{eq_def_fn}) could become problematic. We decide to integrate instead over $\log(\chi)$, since this is the characteristic oscillation length of the factor $\chi^{1-\nu_n} = \cos(\log(\chi)(1-\nu_n))+i\sin(\log(\chi)(1-\nu_n))$.
\dnew
We also implement an algorithm estimating a value $\chi_\mathrm{cut}$ below which 
$\tilde{W}^i(\chi)$ is sufficiently small and contributions to the integral over $\chi$ are negligible. Even if $\chi_\mathrm{cut}$ is  small (e.g. $\chi_\mathrm{cut} \sim 1\text{Mpc}/h\,\, \Rightarrow\,\, z_\mathrm{cut} \sim 2\cdot 10^{-4}$), the lower integration bound in $\log(\chi)$ space is still drastically cut.

\subsection{Behaviour in $k$}\label{sec_behaviour_k}

The transfer functions $T_x(k,\chi)$ of the various contributions to number count and lensing power spectra involve the gauge-invariant perturbations $D$, $\Theta$, $\Phi$ and $\Psi$ defined in \cite{DiDio:2013bqa} and summarized in Appendix \ref{ap_types}, table \ref{tab_gauges}. These account respectively for density, velocity divergence and metric fluctuations. They appear in the transfer functions in combination with various powers of $k$. However, after performing the transformation of section \ref{sec_derivs} in order to eliminate derivatives of Bessel functions, one is left only with contributions from $D$, $\Theta/k^2$, $\Phi$, $\Psi$, $\Phi'$ and $\Psi'$. 
\dnew
The gauge-invariant variables $D$, $\Theta$, $\Phi$ and $\Psi$ behave roughly like the perturbations $\delta_\mathrm{m}$, $\theta_\mathrm{m}$, $\phi$ and $\psi$ of the Newtonian gauge (where the index m stands for non-relativistic matter). It is easy to show with analytical arguments or by looking at the output of Boltzmann codes that the following quantities share roughly the same $k$-dependence, at least at low redshift: $D/k^2 \propto \Theta/k^2 \propto \Phi \propto \Psi \propto \Phi' \propto \Psi'$. This implies that all the transfer functions involved in the problem scale in roughly the same way except for the density one: $T_D(k,\chi)/k^2 \propto  T_{\Theta/k^2}(k,\chi) \propto T_\Phi(k,\chi) \propto \dotsb$.
\dnew
In section~\ref{sec_powerlaw} we have explained how we deal with this situation for non-density terms: instead of performing the FFTlog transformation of $T_x(k,\chi)$, we apply it to $\bar{T}_x(k,\chi) = k^2 T_x(k,\chi)$. This is done at the expense of changing the real part of the complex frequencies $\nu$ for which the integrals $I_\ell(\nu,t)$ must be evaluated. While we need to compute $I_\ell(\nu_n,t)$ always for all $n$'s, for non-density types we need additional computations with $\Re[\nu_n]$ equal to $b$, $b-2$ or $b-4$, with our chosen tilt $b \approx 1.9$.
\dnew
Another approach would be to use a trick from \cite{Assassi:2017lea}, and to transform the density source term $T_D(k,\chi)$ using
\begin{equation}\label{eq_tilt}
k^2 \bessel{\ell}{k \chi} = - \left[\frac{\partial^2}{\partial \chi^2} + \frac{2}{\chi} \frac{\partial}{\partial \chi} - \frac{\ell (\ell+1)}{\chi^2} \right]
\bessel{\ell}{k \chi}~.
\end{equation}
In that way, one can bring all the terms to the same behaviour in $k$, and compute all $I_\ell(\nu_n,t)$ with a unique value of $\Re[\nu_n]$ equal to $b-4$. Since the calculation of the $I_\ell(\nu,t)$ can be done only once and stored for later use, we did not adopt this trick here.
In the last two lines of table \ref{tab_gauges}, we also included the new source terms that need to be computed if one uses equation (\ref{eq_tilt}) for the density. We implemented this method in \texttt{CLASS} for the sake of comparison, and the user can decide to switch to it if desired.

\subsection{Factors of $\ell$}\label{sec_ell}

When considering lensing terms in the calculation of the number count or cosmic shear spectra (or when using the identity (\ref{eq_tilt})),  factors of $\ell$ can appear within the source terms. However, the function $f_n^{ij}(t)$ was designed specifically to be independent of $\ell$. Breaking this independence would theoretically require repeated calculation of the $f_n^{ij}(t)$ for every $\ell$, which we wanted to avoid in the first place. 
\dnew
This is not a problem in practice because the $\ell$-dependence can be factorized for each contribution $C_\ell^{ij, xy}$. 
Noticing that the $\ell$-dependence always appears through factors of $\ell(\ell+1)$ (coming from the angular part of the Laplace operator), we only need to distinguish between three types of contributions $(x,y)$.
\dnew 
If we rename the function which is convolved with $I_\ell(\nu_n,t)$ in the final step of the calculation (equation \ref{eq_def_fn}) as $a_{n,\ell}^{ij}(t)$ instead of $f_n^{ij}(t)$, we see that we can expand this $a_{n,\ell}^{ij}(t)$ as
{\small
\begin{equation}
	a_{n,\ell}^{ij}(t) = \sum_{x,y} f_n^{ij,xy}(t)+ \ell(\ell+1) \left[\sum_{x',y} g_n^{ij,x'y}(t) + \sum_{x,y'} g_n^{ij,xy'}(t)\right] + (\ell(\ell+1))^2 \, \sum_{x',y'} h_n^{ij,x'y'}(t)~,
\end{equation}
}
where the indices without primes run over the source terms with no $\ell$ dependence, and the indices with primes over those with a factor $\ell(\ell+1)$. Then $f$, $g$ and $h$ can all be calculated independently of $\ell$, effectively preserving the separation of cosmology and geometry. This $\ell$-dependence splitting does not require any additional computational time.
	\section{Accuracy and Performance}\label{sec_concl}
	
\subsection{Accuracy}

\begin{figure}[h!]
	\hspace*{-4em} 
	\includegraphics[width=0.55\textwidth]{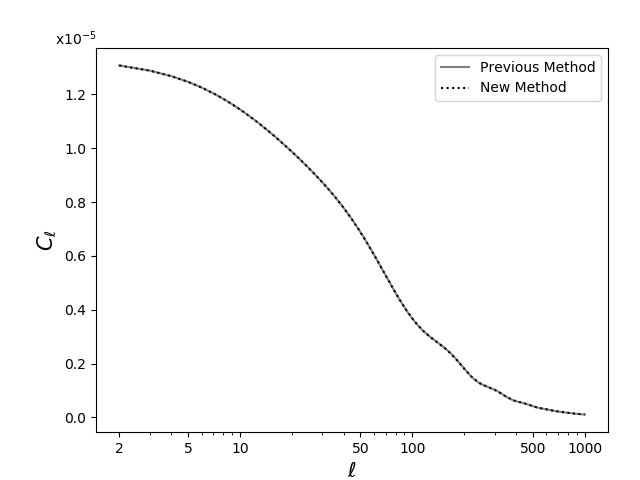}
	\includegraphics[width=0.55\textwidth]{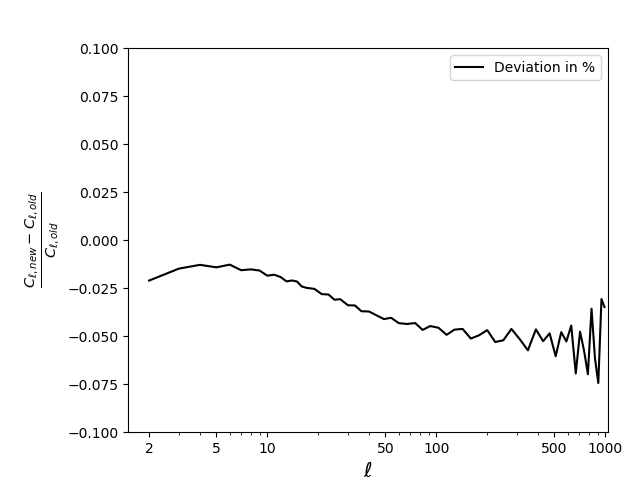}\\
	\hspace*{-4em} 
	\includegraphics[width=0.55\textwidth]{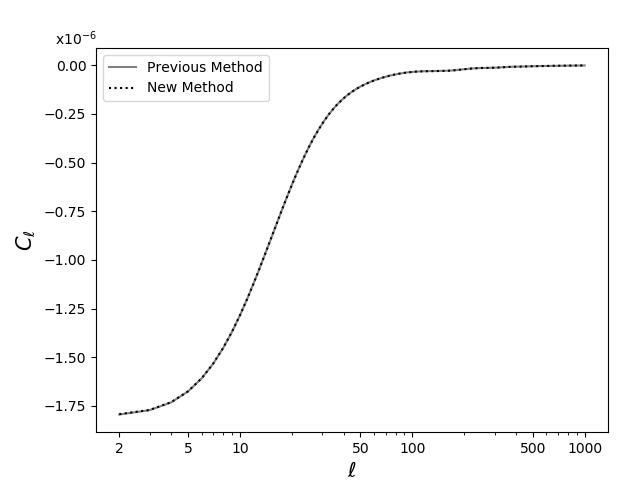}
	\includegraphics[width=0.55\textwidth]{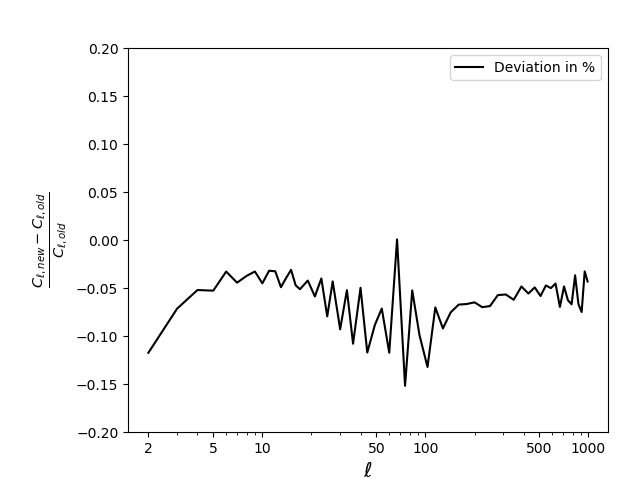}
	\caption{\textit{(Top)} Auto-correlation spectrum of number count (involving all source contributions) in one redshift bin defined by a Gaussian window function with mean redshift $\bar{z}=1.0$ and width $\Delta z = 0.05$\,.
	\textit{(Bottom)} Cross-correlation between two redshift bins defined by two Gaussian windows with $(\bar{z}_1, \Delta z_1) = (1.0, 0.05)$ and $(\bar{z}_2, \Delta z_2) = (1.25, 0.05)$.
	\label{fig_all}}
\end{figure}

\begin{figure}[h!]
	\hspace*{-4em} 
	\includegraphics[width=0.55\textwidth]{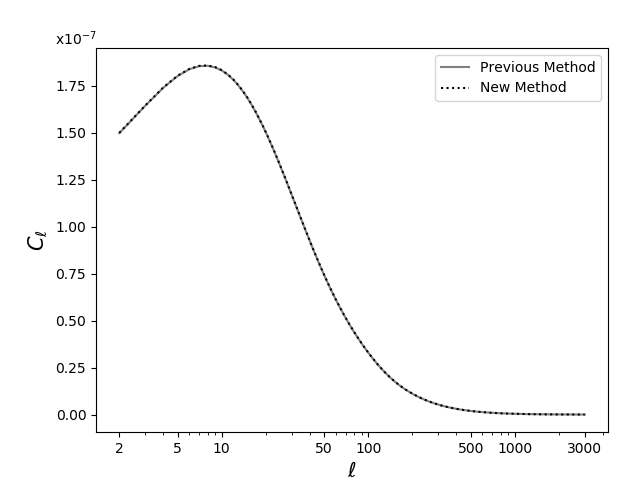}
	\includegraphics[width=0.55\textwidth]{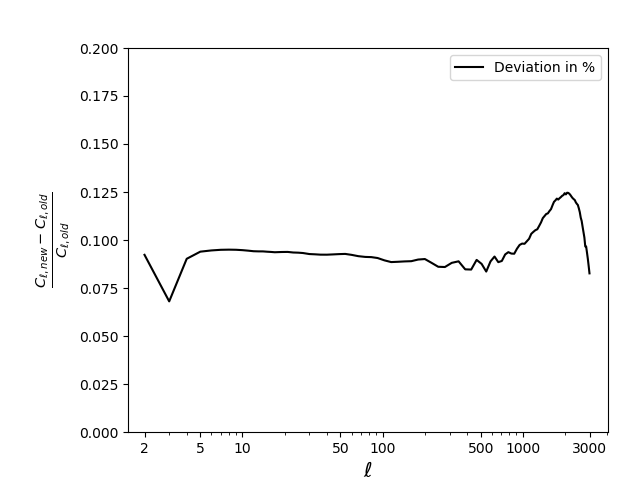}
	\caption{Auto-correlation spectrum of cosmic shear (or more precisely of the lensing potential $C_\ell^{\phi \phi}$) in one redshift bin defined by a Gaussian window function with mean redshift $\bar{z}=1.0$ and width $\Delta z = 0.05$\,.
	\label{fig_shear}}
\end{figure}

\begin{figure}[h!]
	{
	\hspace*{-1em} 
	\includegraphics[width=0.5\textwidth,height=0.45\textwidth]{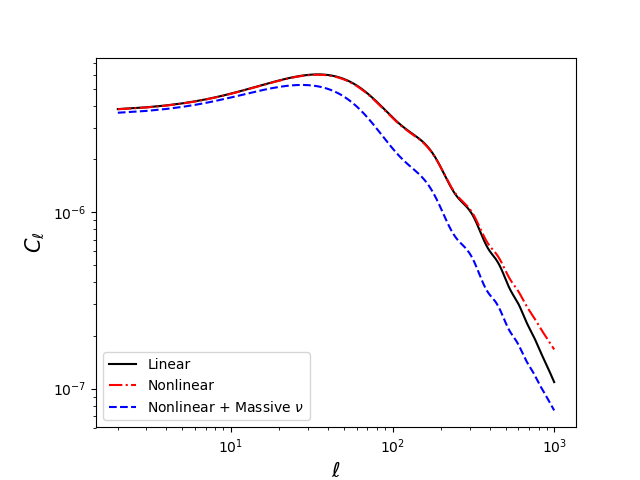} 
	\includegraphics[width=0.5\textwidth,height=0.45\textwidth]{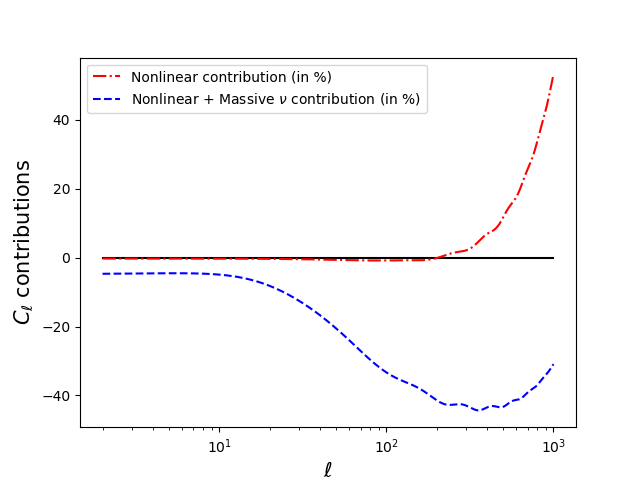}
	}
	\\
	\centering
	\hspace*{-1em}
	\includegraphics[width=0.5\textwidth,height=0.45\textwidth]{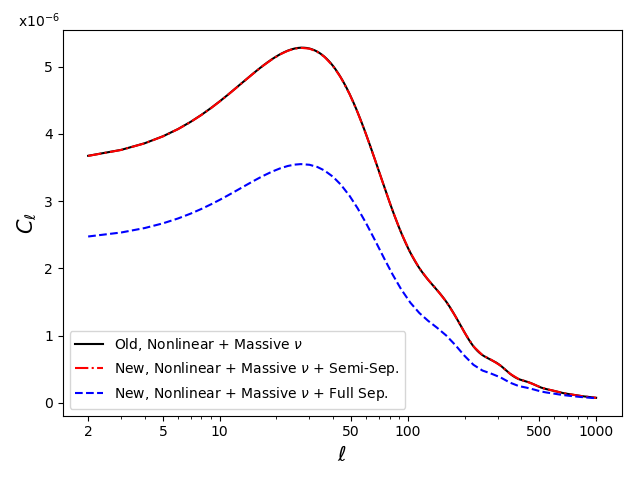}	
	\includegraphics[width=0.5\textwidth,height=0.45\textwidth]{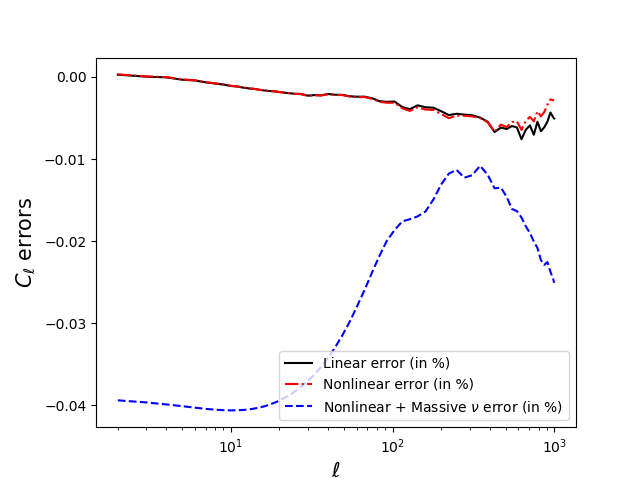}
	\caption{\label{fig_nl_mn}Number count spectra involving only density terms for a redshift bin centered at $\bar{z}=1.0$ with width $\Delta z=0.05$\,.
	\textit{(Top Left)} Total spectra w/o nonlinear corrections from Halofit and massive neutrinos with $M_\nu=1$~eV. 
	\textit{(Top Right)} Impact of these two corrections on the power spectrum, computed as a relative difference (in \%) with respect to the linear spectrum of the massless neutrino model\,.
	\textit{(Bottom left)} Result of the new method with either the \tquote{full separability} or \tquote{semi-separability} approximations compared to the traditional line-of-sight approach.
	\textit{(Bottom right)}  Relative difference (in \%) between the new and old methods. One can immediately see that the additional effects are well captured and the error remains at the sub-permille level.\\
	}
\end{figure}

We first check whether the default \texttt{CLASS} implementation and the new method converge to the same results in the limit of high accuracy settings. We increased the most relevant precision parameters up to the point of saturating the memory limits of a 16-dual-core workstation with 32GB of RAM. In figure \ref{fig_all}, we can see that the agreement on the $C_\ell$'s is in the $0.1\%$\, range for the number count spectra (when all terms and GR corrections are taken into account). The same applies to the cosmic shear $C_\ell$'s, shown in figure \ref{fig_shear}. When only the density source terms are involved, the level of agreement further improves to about $0.01\%$\,(figure \ref{fig_dens}). See table \ref{tab_cosmo} in appendix \ref{sec_further} for the cosmological parameters used for the comparisons.
\begin{figure}[h!]
	\hspace*{-4em} 
	\includegraphics[width=0.55\textwidth]{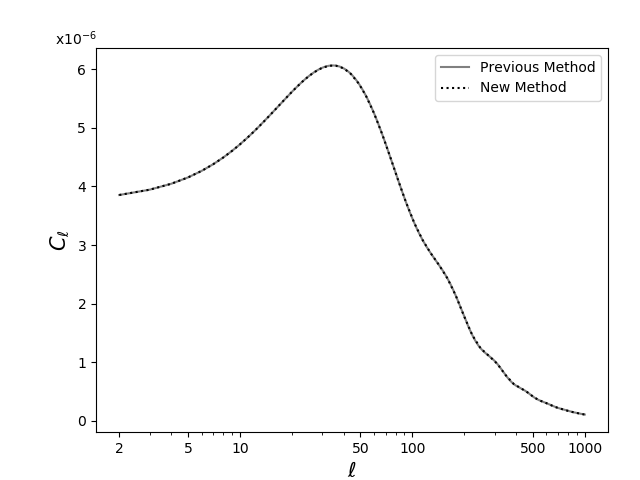}
	\includegraphics[width=0.55\textwidth]{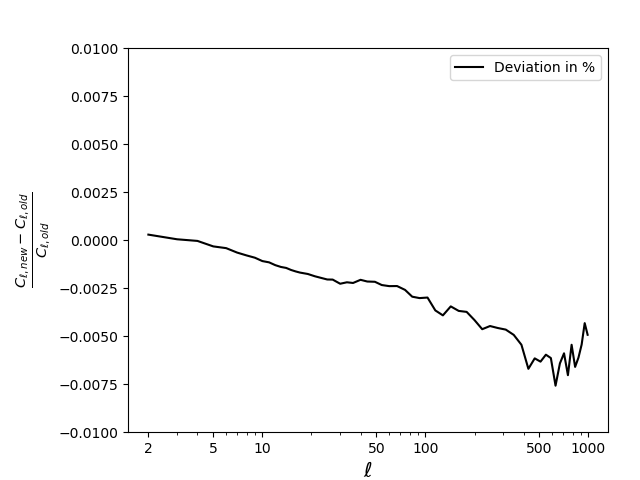}
	\caption{Auto-correlation spectrum of number count (involving only the density source term) in one redshift bin defined by a Gaussian window function with mean redshift $\bar{z}=1.0$ and width $\Delta z = 0.05$\,.
	\label{fig_dens}}
\end{figure}
\noindent These results validate the accuracy of the new method, since an error level of $0.1\%$ on the full number count or cosmic shear spectra is sufficient for fitting the experimental results of future surveys. 
In order to test the \tquote{semi-separable} approximation implemented in our code, we must repeat this exercise in presence of nonlinear corrections and/or massive neutrinos, which both introduce a scale dependence in the density fluctuation growth factor. To this end, we switch on Halofit corrections~\cite{Smith:2002dz,Takahashi:2012em,Bird:2011rb} and/or degenerate massive neutrinos with a total mass $M_\nu=1$~eV. 
This mass choice is rather extreme given that current cosmological upper bounds are in the ballpark of $M_\nu \sim 0.1$~eV to $0.3$~eV, and as such, any lower mass should be captured even better. The effect of these ingredients at redshift $\bar{z}=1$ is visible in the upper panels of figure~\ref{fig_nl_mn}. 
\dnew %
We first check that the \tquote{full separability} approximation would introduce large errors. 
The \tquote{full separability} approximation is implemented in the code as follows: instead of performing the FFTlog transformation at each time step, we only do it once at redshift zero to derive the $c_n$'s of equation (\ref{eq_separability}). We then rescale them at each time using the effective growth factor defined in section~\ref{sec_derivs}. Unsurprisingly, we obtain in this case a large error in the number count spectrum, of the order of 50\% (bottom left panel in figure~\ref{fig_nl_mn}). 
\dnew %
However, our \tquote{semi-separability} method brings the result back in very good agreement with the traditional line-of-sight result. The residual difference between the old and new method is better seen in the bottom right panel of figure \ref{fig_nl_mn}, in the case of number count with only density terms. Adding non-linear corrections leaves the residual at a level below 0.01\%, while switching on a total neutrino mass $M_\nu=1$~eV raises it to around 0.04\%. When considering number count spectra with all GR terms, we find that adding non-linear corrections and/or massive neutrinos keeps the difference at the level of 0.1\%, which is sufficient for future surveys.
\dnew %
Thus the new method --- including all the choices and approximations presented in the previous sections --- is sufficiently accurate. Still, it would be very interesting to investigate what dominates the residual $\sim 0.1$\% differences found for all the spectra including a lensing effect. If there is a clear reason to believe that most of it comes from errors in the traditional line-of-sight approach (as implemented in the default \texttt{CLASS} code), then the new method could actually be significantly more precise than $0.1\%$. Indeed, we believe that this the case, because when we increase the precision settings, the lensing spectrum obtained with the old \texttt{CLASS} code converges very slowly towards a stable result. It is not even fully converged at the sub-percent level when the memory limit is reached. This is not the case with the new method, which already gives stable results for nearly the same precision settings in the case of density and lensing contributions, while using a small amount of memory. The calculation of the density contribution with the old code is also much more stable than that of lensing. Thus it seems that the true level of general agreement between the old and new method might be $0.01\%$ rather than $0.1\%$ due to the fact that the old method has a specific problem to compute lensing spectra with the very high level accuracy.
\dnew
This problematic feature of the default \texttt{CLASS} implementation can be easily explained. It is related to the fact that in the lensing case, the support of the window function $\tilde{W}^i(\chi)$ defined in equation (\ref{eq_lens_window}) reaches $\xmin=0$. Indeed, in principle, the lensing of a source standing at some finite redshift has contributions from modes located arbitrarily close to us, for which a given angle corresponds to infinitely small wavelengths and thus infinitely large wavenumbers. This means that in the  traditional line-of-sight approach, the harmonic transfer functions $\Delta_\ell^{\alpha,i}(k)$ of lensing converge very slowly as a function of $k$. Mathematically this is seen by noticing that these transfer functions are obtained by integrating the product $\tilde{W}^i(\chi) T_\alpha(k,\chi) j_\ell(k \chi)$ over $\chi$. For arbitrarily large $k$, there is always a non-vanishing contribution to the integral from a distance of $\chi \sim \ell/k$. The convergence of $\Delta_\ell^{\alpha,i}(k)$ is only ensured by the fact that the lensing transfer function $T^{\phi+\psi}(k,\chi)$ scales like $k^{-2}$ in the large $k$ limit. This is very different from non-integrated cases, for which convergence arises much earlier simply because the lower edge $\chi^i_\mathrm{min}$ of the support of $W^i(\chi)$ forces $\Delta_\ell^{\alpha,i}(k)$ to vanish above $k \sim \ell/\chi^i_\mathrm{min}$. This slow convergence of $\Delta_\ell^{\alpha,i}(k)$ for lensing term is indeed problematic because the period of oscillation of  $j_\ell(k \chi)$ as a function of $\chi$ is given by $2 \pi / k$. Thus, in order to get precise results, one needs to increase a lot the number of sampled values in the integral over the line-of-sight. This leads to a saturation of the memory before obtaining a sampling that would ensure the convergence of the $C_\ell$'s at the 0.1\% level.
\dnew 
The new method avoids such problems because the integral over slowly-damped oscillating functions at large $k$ is done analytically. The integral over large $k$ is performed within the FFTlog transformation, which does not involve any Bessel function and converges without problems. 
There are no issues of divergences for small $\chi$ either, since the product $\tilde{W}(z) \chi^{1-\nu_n}$ always approaches zero in the limit $\chi\to 0$ for valid choices of $\Re[\nu_n]$ for which the Bessel integral $I_\ell(\nu_n,t)$ is well behaved (see section \ref{sec_powerlaw}). Thus, the lensing spectra can be computed as accurately as the density spectra without requiring significantly more memory. Besides any considerations based on performance, this is a true advantage of the FFTlog method.
\subsection{Performance}\label{sec_performance}
\textit{Precision measure.} We will now estimate the CPU time needed to reach a precision of \mbox{$Q=0.1\%$}, where we define the precision measure as
\begin{equation}
	Q = \sqrt{\frac{1}{N_\ell} \sum_\ell \left(\frac{C_\ell-C_\ell^{ref}}{C_\ell^{ref}}\right)^2}~,
\end{equation}
with $N_\ell$ discrete samples in $\ell$. \dnew To give a more fair comparison, the reference spectra are calculated for each method separately. This means that we disregard in these tests the 0.1\% level residuals between the reference spectra of the two methods. Indeed, as explained in the previous section, we have strong hints that these residuals come from issues of insufficient sampling in the traditional method, but in absence of a definite proof, we do not want to bias our conclusions by this assumption. Thus our analysis quantifies the numerical error coming from the degradation of the precision settings in each method, relative to the reference spectra of the same method.
\dnew
\noindent \textit{$\ell$ sampling.} The default \texttt{CLASS} implementation does not compute every value of $\ell$ separately, but interpolates with a cubic spline interpolation algorithm between a reasonably chosen set of discrete $\ell$ values at which the $C_\ell$'s are actually calculated. We adopt the same strategy in the new algorithm. The list of $\ell$ values at which the $C_\ell$'s are calculated by default (and in our tests) can be found in appendix \ref{sec_further}. Of course, the precision measure $Q$ takes only these $\ell$ values into account. 
\dnew
\noindent \textit{Approximations.} All our performance calculations use the \tquote{full separability}  approximation, but we avoid using the Limber approximation, since we are interested also in effects for small $\ell$ for which the Limber approximation deviates most from the correct result. We looked at three important test cases: the calculation of cosmic shear spectra, of number count spectra involving only density contributions, and of number count spectra involving all contributions, except for the small gravitational corrections labeled as G1-G5 here and in \cite{DiDio:2013bqa}. These are known to be time-consuming but mostly negligible in the final results.
\dnew
\noindent \textit{Setup.} The computation time for the geometrical integrals $I_\ell(\nu,t)$ is explicitly taken out from the direct comparison, since these could be stored as small binary files on the computer, and thus would not need to be calculated every time. The execution times correspond to a single core and thread on an Intel i5-6200U Quadcore (2.3GHz).
\newline
\begin{sloppypar}
\noindent \textit{Precision parameters.} In each test, all relevant precision parameters in the previous \texttt{CLASS} method and the new method are tuned to achieve maximum speed while remaining at the desired accuracy of $Q\simeq0.1$ The tuned parameters of the old method include \texttt{hyper\_sampling\_flat}, guiding the sampling of the spherical Bessel functions, \texttt{selection\_sampling\_bessel} giving the $\chi$ sampling, and \texttt{q\_linstep} and \texttt{q\_logstep\_spline}, both guiding the $k$ sampling. As described in section \ref{sec_cosm_geom}, the cosmological function $f_n^{ij}(t)$ is calculated in a rough grid and interpolated for the final integration. Thus, the tuned parameters of the new method include the number of FFT coefficients $N_c$, the coarse number of $t$ values $N_{t,spline}$, and the number of $t$ values for the final integration $N_t$. In the new method, it is possible to use different $\chi$ samplings for integrated or normal effects, leading to the two additional parameters $N_{\chi, integrated}$ and $N_{\chi, normal}$. The values of the parameters used for the tests can be found in table \ref{tab_params}. 
\end{sloppypar}
{$\;$}
\newline
\noindent\textit{Nomenclature.} In the two methods, we call $N_z$ the number of redshift bins of the survey, and $N_{tot}$ the total number of redshift bin combinations. When considering contributions from all cross correlations between redshift bins, this number is $N_{tot}=N_z(N_z+1)/2$ and grows quadratically. If we include only the first $M$ closest neighbours, the number is instead given by $N_{tot}=(M+1) N_z-M(M+1)/2$ and grows linearly. 
In our tests, for the cosmic shear spectra, we calculated all possible cross correlations between different redshift bins. For the number count spectra, we limit ourselves to the first two neighboring bins ($M=2$), because other correlations are expected to be strongly noise-dominated in future galaxy redshift surveys.

\begin{table}[h!]
	\centering
	\begin{tabular}{|c||c c c|} 
		\hline
		& Cosmic Shear & NC, density only & NC, all (no gr)\Tstrut{2.5}\\ \hline \hline
		Ranges & & & \\ \hline
		$\ell_{max}$ & 3000 & 1000 & 1000 \Tstrut{2.5}\\
		Number of $\ell$ & 112 & 62 & 62 \\
		Non-Diagonals & All & 2 & 2 \\
		$\Delta z$ & 0.05 & 0.01 & 0.01 \\
		\hline \hline
		Old parameters& & & \\ \hline
		{\tt hyper\_sampling\_flat} &  7.5 & 7 & 8 \Tstrut{2.5} \\
		{\tt selection\_sampling\_bessel} &  1.5 & 0.9 & 1.7 \Tstrut{2} \\
		{\tt q\_logstep\_spline} &  120 & 21 & 21 \Tstrut{2} \\
		{\tt q\_linstep} (ignored)& 100000 & 100000 & 100000 \Tstrut{2} \\
		\hline
		\hline
		New parameters & & & \\ \hline
		$N_{\chi, integrated}$ & 75 & - & 50 \Tstrut{2.5}\\
		$N_{\chi, normal}$ & - & 15 & 25 \\
		$N_{t, spline}$ & 35 & 20 & 70 \\
		$N_t$ & 130 &50 & 100\\
		$N_c$ & 100 & 95 & 95 \\ \hline
	\end{tabular}
	\caption{\label{tab_params}Parameter values for the different timing tests. NC = Number count. The parameter {\tt q\_linstep} is relevant for small wavenumbers and thus mainly for CMB observables: in our case it generally did not affect the results.}
\end{table}

\FloatBarrier
\begin{table}[h!]
	\centering
	\begin{tabular}{|c c||c|c||c|c||c|}
		\hline 
		& & \multicolumn{2}{c||}{Old} & \multicolumn{2}{c||}{New} & \Tstrut{2.5} \\
		$N_{tot}$ &$N_z$ & \quad $T$ [s] \enskip\enskip & $T$/$N_z$ [s]& \quad $T$ [s] \enskip\enskip & $T$/$N_ {tot}$ [ms] \enskip & \enskip $I_\ell(\nu,t)$ [s] \Tstrut{2} \\
		\hline
		5050 &100 & 850 &  8.5 & 20 & 4.0& 2.2\Tstrut{2.5}\\
		2556 &71 & 620 &  8.7  & 11 & 4.4 & 2.6\\
		1326 &51 & 450 &  8.8& 6.1 & 4.6 & 2.5 \\
		351 &26 & 240 & 9.2 & 1.6 & 4.6 & 2.6\\
		105 &14 & 130 & 9.4 & 0.50 & 4.8 & 2.7\\
		10 & 4 & 42 & 10.5 & 0.12 & 12 & 2.1\\ 
		\hline
	\end{tabular}
	\caption{\label{tab_shear}Cosmic Shear. A comparison of the computation times $T$ for each method, revealing that the old method scales as $N_z$, while the new method scales as $N_{tot}$. The new method is up to 350 times faster. The last column gives the time needed to calculate the $I_\ell(\nu,t)$ integrals, which is of minor importance, given that this calculation can be done once and for all.}
\end{table}
\FloatBarrier
\begin{table}[h!]
	\centering
	\begin{tabular}{|c c||c|c||c|c||c|}
		\hline
		& & \multicolumn{2}{c||}{Old} & \multicolumn{2}{c||}{New} & \Tstrut{2.5}\\
		$N_{tot}$ &$N_z$ & \quad $T$ [s] \enskip\enskip & $T$/$N_z$ [s]& \quad $T$ [s] \enskip\enskip & $T$/$N_ {tot}$ [ms] \enskip & \enskip $I_\ell(\nu,t)$ [s] \Tstrut{2} \\
		\hline
		297 & 100 & 13 & 130 & 5.5 & 19 & 2.4 \Tstrut{2.5}\\
		210 & 71 &  9.3 &  130  & 3.8 & 18 & 2.5\\
		150 & 51 & 6.7 & 130 & 2.8 & 19 & 2.5\\
		75 & 26 & 3.4 & 130 & 1.2 & 16 & 2.5\\
		39 & 14 & 2.0 & 140 & 0.60 & 15 & 2.5\\
		9 & 4 & 0.60 & 150 & 0.24 & 27 & 2.1\\ 
		\hline
	\end{tabular}
	\caption{\label{tab_nc}Number Count, density contributions only. A comparison of the computation times $T$ for each method, revealing that the old method scales as $N_z$, while the new method scales as $N_{tot}$. The new method is up to 2.5 times faster.}
\end{table}
\FloatBarrier
\begin{table}[h!]
	\centering
	\begin{tabular}{|c c||c|c||c|c||c|}
		\hline
		& & \multicolumn{2}{c||}{Old} & \multicolumn{2}{c||}{New} & \Tstrut{2.5}\\
		$N_{tot}$ &$N_z$ & \quad $T$ [s] \enskip\enskip & $T$/$N_z$ [s]& \quad $T$ [s] \enskip\enskip & $T$/$N_ {tot}$ [ms] \enskip & \enskip $I_\ell(\nu,t)$ [s] \Tstrut{2} \\
		\hline
		297 & 100 & 7500 & 75& 27 & 90 & 7.8 \Tstrut{2.5}\\
		210 & 71 &  5400 &  76  & 19 & 90 & 8.0\\
		150 & 51 & 3800 & 74 & 14 & 93 & 7.8\\
		75 & 26 & 2000 & 77 & 7.2 & 96 & 7.9\\
		39 & 14 & 1100 & 79 & 3.2 & 82 & 7.9\\
		9 & 4 & 300 & 75 & 0.80 & 89 & 7.9\\ 
		\hline
	\end{tabular}
	\caption{\label{tab_nc_all}Number Count, all contributions except for gravitational terms. A comparison of the computation times $T$ for each method, revealing that the old method scales as $N_z$, while the new method scales as $N_{tot}$. The new method is up to 380 times faster.}
	\vspace*{-1em}
\end{table}
\FloatBarrier
{$\;$}
\dnew 
\textit{Speedup.} The performance comparison between the two methods for different numbers of observable redshift bins is given in Tables \ref{tab_shear}, \ref{tab_nc}, and \ref{tab_nc_all}. The new method is always faster in our tests. The speedup of the new method ranges in our tests from factors of around 2 (density only) to 400 (cosmic shear or lensed density). As mentioned in section \ref{sec_behaviour_k}, if multiple sources are calculated, multiple $\Re[\nu]$ are required for the $I_\ell(\nu,t)$, explaining the slightly longer times in the last column of Table~\ref{tab_nc_all} compared those in Tables~\ref{tab_shear} and \ref{tab_nc} ($8$s instead of $2.5$s). This would not be the case if we used the \tquote{tilt reduction} method of section \ref{sec_powerlaw}, but this issue is anyway unimportant, give that the $I_\ell(\nu,t)$'s calculation can be done once and for all. The results can be stored in a binary file and quickly retrieved in later executions.
\dnew
\textit{Scaling with $N_z$ and $N_{tot}$.} The new method scales with $N_{tot}$ instead of $N_z$ (like the old one) due to the different arrangement of integrations. Thus, when we consider cosmic shear with all cross-correlation spectra, the new method scales quadratically with $N_z$, while the old one scales linearly.  This could be a problem for the new method when many bins are considered, but fortunately the computation time per spectrum is much smaller in this method, of the order of $T/N_{tot} \simeq 5\,$ms, instead of $T/N_{z} \simeq 9\,$s in the old method for the same accuracy. Thus the new method is about 350 faster for 4 bins, 42 times faster for one hundred bins, and the old method would only overtake for $\sim$ 4000 redshift bins (8 million total combinations). One should keep in mind that cosmic shear analyses will always be optimal when analysed with a rather small number of redshift bins, in order to get good statistics in each bin: thus the most interesting cases are the ones with a dozen of bins at most, for which we find the best gains with the new method.
\\
When considering instead cluster counts with only two non-diagonal terms ($M=2$), both method scale linearly with $N_z$, and the gain is roughly independent of the number of bins. This gain is still very large when including integrated terms like lensing, since the new method gives accurate spectra without requiring integrals over slowly-converging oscillatory functions in $k$ space. We find a speed up by a factor of the order of 380 in that case. When including only the leading density term, the new method is still faster, by about 2.5. This comparison is valid for $M=2$: the new method performs even better when considering $M=1$ or when calculating only auto-correlation spectra, while the old method would overtake for $M\geq8$ when computing density terms only. 
\dnew
\textit{Scaling with precision parameters.} We have also qualitatively evaluated how the new method scales with the different precision parameters. Most scaling relations are approximately linear, although with some minimal constant offset. This is likely due to other parts of the code becoming the dominant time consuming contributions in those cases. The parameters $N_c$, $N_{\chi}$, $N_t$, $N_{t,spline}$\,, and $N_\ell$ show this scaling. The calculation of the geometrical integrals $I_\ell(\nu,t)$ on the other hand is linearly proportional to $N_c$, $\ell_{max}$\,, and the number of $t$ values for which they are calculated. It also depends non-trivially on $Re[\nu]$\,.
\dnew 
\textit{Scaling with bin width.} In the old method, the $\chi$ integration over a very narrow redshift bin around some $\chi^{*}$ results in a particular $\bessel{\ell}{k \chi^{*}}$ effectively being selected, thus the $k$ sampling has to be rather precise to effectively capture the oscillatory nature of the spherical Bessel function. Schematically, we can describe this by approximating the narrow Gaussian window function as a Dirac-delta distribution,
\begin{equation}
\begin{split}
\Delta^{\alpha,i}(k) = & \intinf d\chi \; T_\alpha(k,\chi) \; \bessel{\ell}{k \chi} \; W^i(\chi) \\ \approx &\intinf d\chi \; T_\alpha(k,\chi) \; \bessel{\ell}{k \chi} \; \delta(\chi-\chi^{*}) \\ = & \;T_\alpha (k,\chi^{*}) \; \bessel{\ell}{k \chi^{*}} ~.
\end{split}
\end{equation}
Thus the final integral in $k$ oscillates like the Spherical Bessel function, meaning it has to be evaluated with a rather fine grid. A very broad redshift bin instead effectively results in averaging the $\bessel{\ell}{k \chi}$ over several oscillations, canceling out some of the oscillatory nature of the Bessel functions. Of course a broader redshift window requires more sampling points in $\chi$ instead. Thus there is a trade-off, and for number count the old method is quicker for intermediate bin widths ($\Delta z \sim 0.01$) than for very wide or thin bins. For the new method, the Bessel functions are already integrated to obtain $I_\ell(\nu,t)$\,, always canceling out most of the oscillations. However, as the window width is shrunk, a smaller range of $t$ values is allowed, leading to a speedup of the new method. Table~\ref{tab_width} shows that the new method is always faster, but the gain is minimal for $\Delta z \sim 0.01$, and grows on both sides of this value. Note that the previous performance tests of  this section assumed with $\Delta z \sim 0.01$, which is the most unfavorable case for the new method, with a factor two improvement for number count with density terms only. With $\Delta z \sim 0.001$ the speed up factor reaches 40. Thus the new method will perform very well for spectroscopic surveys with typical bin widths $\Delta z \sim 0.001(1+z)$.
\begin{table}[h!]
	\centering
	\begin{tabular}{|c c||c c c c c||c c c c c | c |}
		\hline
	 & & \multicolumn{5}{c||}{Old}  & \multicolumn{5}{c |}{New (without $I_\ell(\nu,t)$)}   & \Tstrut{2.5}  \\
	 &$N_z$ & 100 & 71 & 51 & 26 & 14  & 100 & 71 & 51 & 26 &14 \Tstrut{3} & \\
	 Width & & & & & & & & & & & & Speedup\\
		\hline
		0.5 & & 62 & 48 & 33 & 17 & 9.0 & 7.8 & 5.5 & 4.0 & 2.0 & 1.0 \Tstrut{2.5}& 8\\
		0.1 & & 42 & 33 & 22 & 11 & 6 & 8.4 & 6.0 & 4.4 & 2.2 & 1.1\Tstrut{2.5}& 5\\
		0.05 & & 37 & 31 & 20 & 10 & 5.6 & 7.4 & 5.2 & 3.8 & 1.86 & 0.96 & 5\\
		0.01 & & 13 & 9.3 & 6.7 & 3.4 & 2.0  & 5.5 & 3.8 & 2.8 & 1.2 & 0.6 \Tstrut{2.5} & 2\\
		0.005 & & 60 & 48 & 31 & 16 & 8 & 3.6 & 2.5 & 1.8 & 0.86 & 0.41\Tstrut{2.5} & 20\\
		0.001 & & 80 & 45 & 33 & 17 & 9 & 1.9 & 1.3 & 0.93 & 0.45 & 0.22\Tstrut{2.5} & 40\\
		\hline
	\end{tabular}
	\caption{\label{tab_width}The scaling of the different methods with the redshift bin width for number count spectra. While the old method takes longer for either very broad or very thin redshift bins, the new method always increases in speed for thinner bins. The precision parameters for these runs can be found in the appendix, table \ref{tab_width_params}. Our previous tests were conservatively performed in the case $\Delta z = 0.01$ leading to minimal speedup.} \vspace*{-1em}
\end{table}
\FloatBarrier

	\section{Conclusions}
In this paper we have further developed and implemented in the \texttt{CLASS} code a method proposed in \cite{Assassi:2017lea} (see also \cite{Gebhardt:2017chz}) to efficiently evaluate the angular power spectra for large-scale structure observables. We particularly focus on the angular power spectrum of the number counts including all relevant terms such as redshift-space distortions, weak lensing, relativistic corrections etc., and the angular power spectrum of cosmic shear. The method is based on the power-law decomposition of the $k$-dependent function integrated over spherical Bessel functions which allows for the analytical solutions of the momentum integrals. This power-law decomposition is achieved using the FFTlog algorithm. Compared to earlier studies we make one step further in separating the functions that depend on cosmology form cosmology-independent geometrical factors. We also discuss and test many important aspects of practical implementation of the algorithm. All these improvements lead to significant gain in speed and accuracy. 
\dnew
In order to test our code we evaluate the angular power spectra for number counts and cosmic shear for many different configurations of window functions and redshift bins and compare it to the default \texttt{CLASS} implementation. We find that the new method is always more efficient, with the speedup up to a factor of $\mathcal O (400)$. At the same time the accuracy of the results remains very good, with the relative error below $0.1\%$. There are reasons to believe that this error is due to numerical integration in the standard approach and that the accuracy of the new method can be easily an order of magnitude better. Given that the relative error of $0.1\%$ is sufficient for all practical purposes for current and planned large-scale structure surveys, we leave the further investigation of this issue for future work. We plan to publicly release the CLASS modifications discussed in this work within a few months, after further polishing and minor improvements in the new parts of the code.
\dnew 
Another clever way of speeding up Boltzmann codes has been proposed by the authors of \cite{Campagne:2017xps}. This approach, called Angpow, sticks to the same sequence of operations as the traditional line-of-sight approach until the calculation of the transfer functions $\Delta_\ell^{\alpha,i}(k)$. However the latter are sampled in particular $k$ values, allowing to compute the $C_\ell^{\alpha \beta,ij}$ efficiently with a Clenshaw-Curtis-Chebyshev algorithm. This allows to reduce the number of sampled $k$ values by a significant amount. We note that this method is interesting per se, but cannot be combined with the current approach since we infer the $C_\ell^{\alpha \beta,ij}$ through a very different integral.
\dnew 
Throughout this paper we have focused on the power spectrum only and we have discussed only the large-scale structure observables. However, the method we presented is equally applicable to other cases such as for example the CMB primary anisotropies. It would be very interesting to see whether similar improvements are possible for the CMB observables. Another direction for future work is the application of our method to higher point correlation functions. In particular, as shown in \cite{Assassi:2017lea}, the case of the CMB or large-scale structure bispectra seems promising. 
Yet another possible application of our method is efficient evaluation of the covariance matrix for angular power spectra. All these examples are quite relevant and it would be worth exploring them in more details in the future.

	\section{Acknowledgements}
	M.S. gratefully acknowledges support from the Institute for Advanced Study and the Raymond and Beverly Sackler Foundation. M.Z. is supported by NSF grants AST-1409709 and PHY-1521097, the Canadian Institute for Advanced Research (CIFAR) program on Gravity and the Extreme Universe and the Simons Foundation Modern Inflationary Cosmology initiative.
	
 	\clearpage
\appendix
\section{Types of Source Terms}\label{ap_types}
We summarize in Table~\ref{tab_terms} the different source functions contributing to the number count spectra. 
\begin{table}[h!]
	\hspace*{-2em}
	\begin{tabular}{| c | c | c | c | c |}
		\hline 
		Name & Prefactor & Window function & Source & $k$ scaling\Tstrut{2.5}\\[1pt]
		\hline \hline 
		Density &$1$ & $b W$ & $T^D$  & $1$ \Tstrut{2.6}\\ \hline
		Lensing & $\ell(\ell+1)$ & ${\displaystyle \bclap{\int\limits_\chi^\infty} \,\,d\chi'\, \left(\frac{2-5s}{2}\right) \frac{\chi'-\chi}{\chi \, \chi'} \,\, W }$ & $T^{\phi+\psi}$ & $1/k^2$ \Tstrut{5.5} \\[18pt] \hline
		Doppler 1 & ${\displaystyle - \frac{1}{k^2} \frac{\partial}{\partial\chi}}$ & ${\displaystyle W \left(1+\frac{H'}{aH}+\frac{2-5s}{\chi aH} + 5s - f_{evo}\right) }$ & $T^\theta$ & $1/k^2$ \Tstrut{5}\\
		Doppler 2 & ${\displaystyle \frac{1}{k^2}}$ & $W (f_{evo}-3) aH$ & $T^\theta$ & $1/k^2$\Tstrut{3}\\
		RSD & ${\displaystyle\frac{1}{k^2} \frac{\partial^2}{\partial\chi^2}}$ & ${\displaystyle \frac{1}{aH} W}$  & $T^\theta$ & $1/k^2$\Tstrut{4}\\[10pt] \hline
		GR 1 & 1 & ${\displaystyle W \left(2+\frac{H'}{aH^2}+\frac{2-5s}{\chi aH} + 5s - f_{evo}\right)}$ & $T^\psi$ & $1/k^2$\Tstrut{4.5}\\
		GR 2 & 1 & $- W (2-5s) $ & $T^\psi$ & $1/k^2$\Tstrut{3}\\
		GR 3 & 1 & $W / (aH)$ & $T^{\psi'}$ & $1/k^2$\Tstrut{3}\\
		GR 4 & 1 & ${\displaystyle\int\limits_\chi^\infty d\chi' \left(\frac{2-5s}{\chi'}\right)\,\,W}$ & $T^{\phi+\psi} $& $1/k^2$ \Tstrut{6}\\
		GR 5 & 1 & ${\displaystyle \int\limits_\chi^\infty d\chi' \left(1+\frac{H'}{aH}+\frac{2-5s}{\chi' aH}+5s-f_{evo}\right)\,\,W}$ & $T^{(\phi+\psi)'}$& $1/k^2$\Tstrut{5}\\[18pt] \hline \hline
		Density Split 1 & $\mathcal{D}$ & $b W$ & $T^D$ & $1/k^2$ \Tstrut{3}\\
		Density Split 2 & ${\displaystyle \frac{\ell(\ell+1)}{\chi^2}}$ & $b W$ & $T^D$ & $1/k^2$ \Tstrut{4} \\[8pt] \hline 
	\end{tabular}
	\caption{Different terms appearing in the total source functions with their respective prefactors. $b$ is the (linear) bias, $s$ is the magnification bias, $f_{evo}$ is a possible evolution of $dN/dz$, all in accordance with \cite{DiDio:2013bqa}, where they are defined more precisely. The layout of the table is discussed further in the text below.\label{tab_terms}}
\end{table}
\FloatBarrier
\noindent The column \tquote{Window function} corresponds to the combination  $D(\chi) W^i(\chi)$ or $D(\chi) \tilde{W}^i(\chi)$ of the main text (e.g. equation (\ref{eq_WD})\,), and the column \tquote{Source} to what is called $T_\alpha(k,\chi)$. The \tquote{Prefactor} is an overall factor for the window function which comes from the different operations in section \ref{sec_types}, like the integration by parts leading to factors of $\partial/\partial \chi$ in section \ref{sec_derivs}, the $\ell$ factors in section \ref{sec_ell}, and the $k$ behavior intrinsic to the window function as in section \ref{sec_behaviour_k}. For convenience, we also list the resulting overall behavior of the product of source and window functions in the column \tquote{$k$ scaling}.
The source function contributing to the cosmic shear spectra is identical to the \mbox{\tquote{lensing}} source function (second entry of the table).
\dnew %
The transfer functions refer to gauge-independent variables whose expression in the newtonian and synchronous gauge is explicitely given in Table~\ref{tab_gauges} below.
\noindent The last two entries in Table~\ref{tab_terms} are not used in our default implementation. We switch from the \tquote{Density} to  \tquote{Density Split} terms when testing the method that uses equation (\ref{eq_tilt}) to reduce the tilt. These sources use the operator $\mathcal{D}$ given by
\begin{equation}
\mathcal{D} = \left[-\frac{\partial^2}{\partial \chi^2} + \frac{2}{\chi} \frac{\partial}{\partial \chi} - \frac{2}{\chi^2} \right]~.
\end{equation}
In Table \ref{tab_gauges} we summarize the different gauge invariant source functions $T_{\alpha}(k,\chi)$ used throughout this paper.
\begin{table}[h!]
	\centering
	\begin{tabular}{|c|c|c|} 
		\hline
		Source Name \& Symbol & Newtonian Gauge & Synchronous Gauge \\
		\hline
		Density $T^D$& $\delta_m + 3 \frac{a H}{k^2} \theta_m$  & $\delta_m + 3 \frac{a H}{k^2} \theta_m$ \Tstrut{3}\\ 
		Velocity $T^\Theta$& $\theta_m$  & $\theta_m + k^2 \alpha$ \Tstrut{2.5}\\ \hline
		Bardeen Potential $T^\Phi$& $\phi$  & $\eta - \mathcal{H} \alpha$\Tstrut{2.5}\\ 
		Bardeen Potential $T^\Psi$& $\psi$  & $\mathcal{H} \alpha + \alpha'$\Tstrut{2.5}\\ \hline
	\end{tabular}
	\caption{Gauge invariant quantities in Newtonian and Synchronous Gauge. These correspond to the source functions in \texttt{CLASS}. For the notation of Synchronous Gauge see \cite{Ma:1994dv}. See also \cite{DiDio:2013bqa}.\label{tab_gauges}}
\end{table}
\FloatBarrier
\vspace*{-1.5em}
\section{Details on the Power Law Decomposition}\label{ap_powerlaw}
In this section we want to discuss how the power law decomposition in equation (\ref{eq_power_law_expansion}) is done in practice on a finite-precision computer. We want to perform an expansion of the type
\begin{equation} \tag{\ref{eq_power_law_expansion}}
f(\vx; k) = \sumn c_n(\vx) \,\, k^{\nu_n}~, \quad \text{where} \enskip \nu_n = \frac{2\pi i \, n}{\log(\kmax/\kmin)} +b~.
\end{equation} 
First, we should note that the precise form of $\nu_n$ depends on our choice of sampling in $\log(k)$. In our code, we use a slightly modified definition,
\begin{equation}
 \tilde{\nu}_n = \frac{2\pi i \, n}{T}\frac{N-1}{N} +b ~,
\end{equation}
where $N$ is the number of samples in $\log(k)$ and $T=\log(\kmax/\kmin)$. 
\clearpage
\noindent This choice is based on the following argument: For $N$ evenly sampled values of $\log(k_m)$ with $m \in \lbrace 0, \, ...\, , N-1 \rbrace$ we obtain
\begin{equation}
\log(k_m) = \log(\kmin) + \frac{m}{N-1} \left(\log(\kmax)-\log(\kmin)\right)~,
\end{equation}
such that the sampling is linear, starts at \kmin\,and ends at \kmax\,.
\dnew %
If we first set the tilt $b$ to zero, the FFT coefficients calculated by the direct FFT algorithm are simply given as
\begin{equation}
\tilde{c}_n = \summ f(k_m) \exp(-2 \pi i \, n \, m /N)~,
\end{equation}
and transforming them according to 
\begin{equation}
c_n = \frac{1}{N} (\kmin)^{-\tilde{\nu}_n} \tilde{c}_n = \frac{1}{N} \exp\left(-\log(\kmin) 2 \pi i  \, n/T \frac{N-1}{N} \right) \tilde{c}_n~,
\end{equation}
 gives us the coefficients with our desired back-transformation properties
\begin{equation} \label{ap_eq_fftproof}
\begin{split}
\sumn c_n k_a^{\tilde{\nu}_n} & =\summ\sumn f(k_m) \exp(-2 \pi i n m /N) \exp\left(-\log(\kmin) 2 \pi i  \, n/T \frac{N-1}{N} \right)k_a^{\tilde{\nu}_n} /N\\
& =\summ\sumn f(k_m) \exp(-2 \pi i \, n \, m /N) \exp\left((\log(k_a)-\log(\kmin)) 2 \pi i  \, n/T \frac{N-1}{N} \right) /N \\
& =\summ\sumn f(k_m) \exp(-2 \pi i \, n \, m /N) \exp\left( \frac{a}{N-1} 2 \pi i  \, n \frac{N-1}{N} \right) /N \\
& =\summ\sumn f(k_m) \exp(2 \pi \, i \, n (a-m) n/N)/N\\
& =\summ f(k_m) \delta_{am} = f(k_a) ~.
\end{split}
\end{equation}
The factor $(N-1)/N$ in the definition of the $\tilde{\nu}_n$ can thus be explicitly seen as a \tquote{correction} for our choice of sampling in $\log(k)$, which involved a scaling $m/(N-1)$ instead of $m/N$ for the usual FFT. This sampling was chosen in such a way as to have \kmax\, be the maximum sampled value.
\dnew
Transforming instead $f(k)k^{-b}$ introduces the tilt $b$ in our definition of the Fourier mode $\nu_n$. If we choose to transform precisely $f(k)(k/\kmin)^{-b}$, the relation $c_n = \frac{1}{N} (\kmin)^{-\tilde{\nu}_n} \tilde{c}_n$ remains valid even for general $b\neq 0$\,. Overall, the coefficients $c_n$ can be obtained by a fast implementation using the FFT algorithm and a simple multiplication by $(\kmin)^{-\tilde{\nu}_n}/N$.

\clearpage
\section{Transformations of the Bessel Integrals} \label{ap_traffo}

\begin{figure}[h!]
	\includegraphics[width=0.45\textwidth]{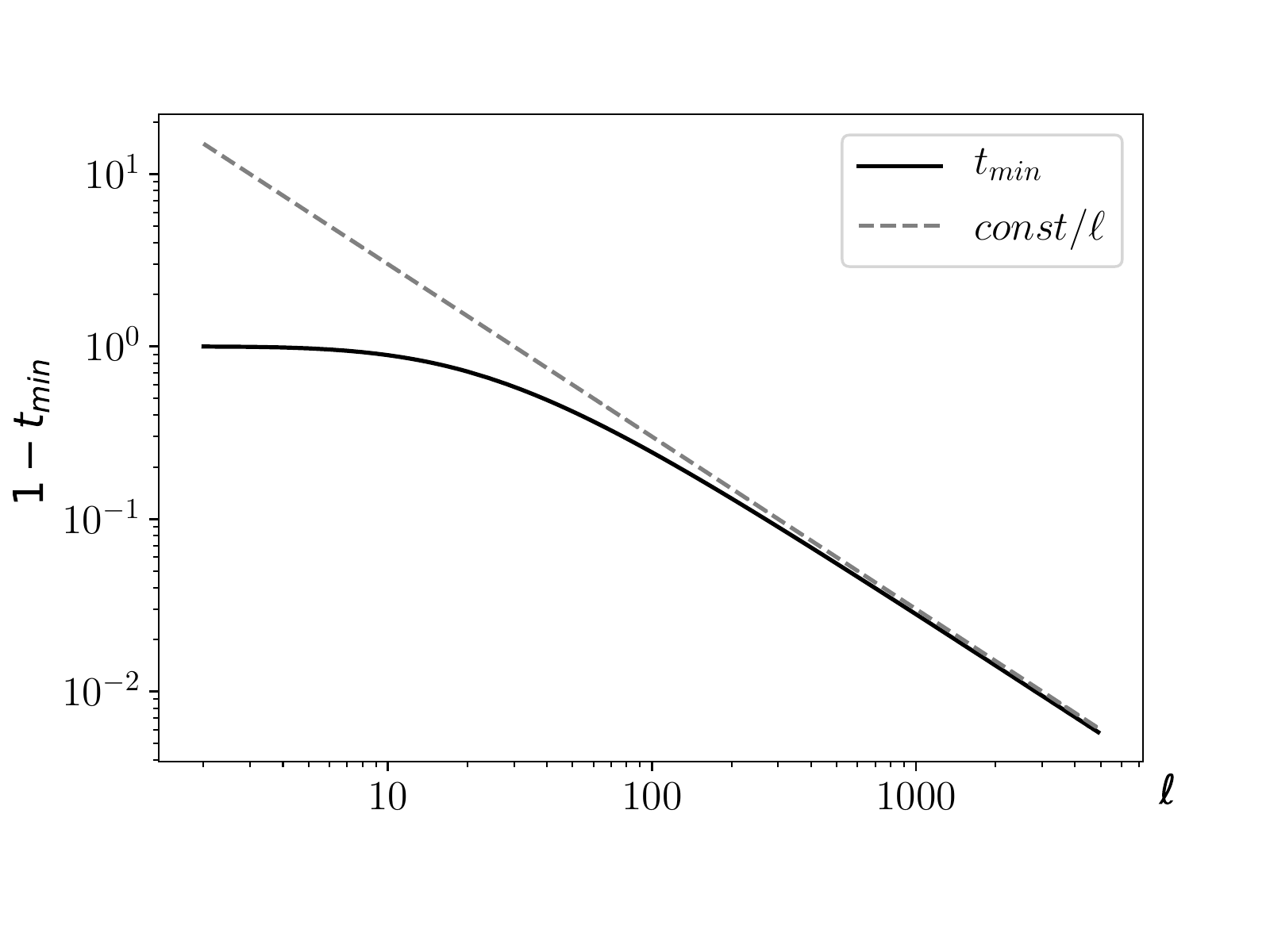}
	\includegraphics[width=0.45\textwidth]{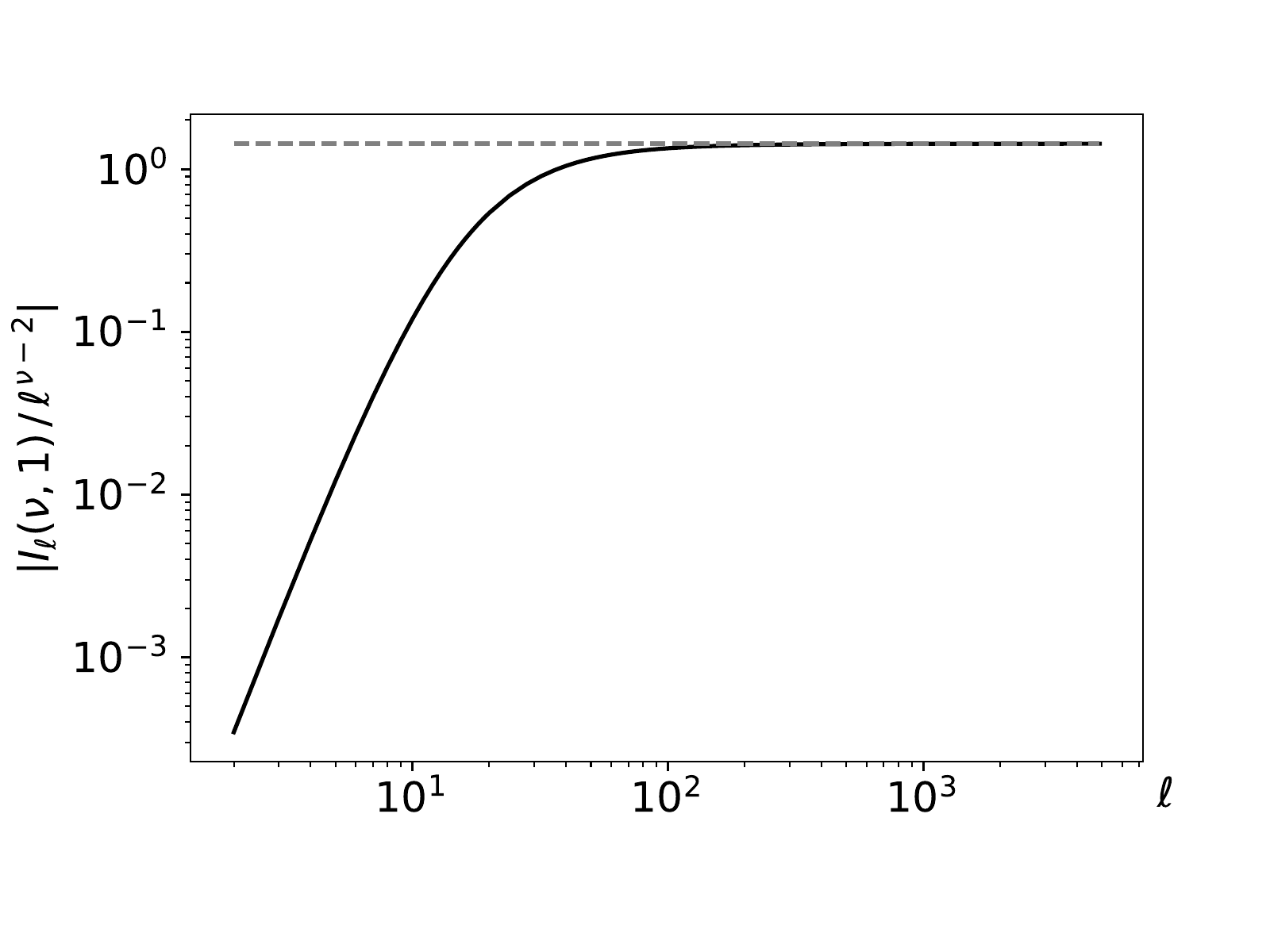}
	\caption{Another consequence of the Limber limit: For large $\ell$ the $t_{min}$ parameter behaves as $1/\ell$ (left), and the $|I_\ell(\nu,1)/\ell^{\nu-2}|$ is constant as in equation \ref{eq_Il_limit} (right). Note that the oscillations due to imaginary $\nu$ are correctly captured and the relative size approaches the correct constant. The black lines indicate the behavior for $\nu=-2.1+30i$ and $\epsilon=10^{-4}$, while the grey lines specify asymptotes. On the left, the grey line is $\ell^{-1}$ times an arbitrary constant (here $30/\ell$), while on the right side the constant is fixed by \ref{eq_Il_limit}. The constant for $t_{min}$ is not exactly $\log(1/\epsilon)$ because of the influence of the hypergeometric function.\label{fig_tmin}}
\end{figure}

\textit{Analytic limit.} Many properties used in the following section can be found on \\
\mbox{\url{http://functions.wolfram.com/HypergeometricFunctions/Hypergeometric2F1}}\,.\\ Further discussion on hypergeometric functions can also be found in \cite{Assassi:2017lea}~.\dnew 
First we note that 
\begin{equation}
	\hyp{a}{b}{c}{1} = \frac{\Gamma(c)\Gamma(c-a-b)}{\Gamma(c-a)\Gamma(c-b)}~, \qquad \text{where} \quad {\rm Re}[c-a-b]>0 ~.
\end{equation}
In our case ${\rm Re}[c-a-b] = 2-{\rm Re}[\nu_n]$, requiring ${\rm Re}[\nu_n]<2$~. This corresponds to the analytical result 
\begin{equation}
	I_\ell(\nu,1) = \frac{\pi^{\frac{3}{2}}\Gamma(\ell+\frac{\nu}{2})\Gamma(1-\frac{\nu}{2})}{ \Gamma(\frac{3-\nu}{2})\Gamma(\ell+2-\frac{\nu}{2})}~.
\end{equation}
This is the analytic limit for $t\to 1$. 
\dnew
\textit{Limber limit.} We now want to discuss the Limber limit, not used in our implementation but mentioned in equation (\ref{eq_limber}) for self-consistency checks. First, we want to note that one has to be careful not to double-count the point $t=1$ when using equation (\ref{eq_limber}). Thus, the replacement for the $I_\ell(\nu,t)$ for the integration from $t$ in $[0,1]$ is actually 
\begin{equation}
	I_\ell(\nu,t)\xrightarrow[\ell \to \infty]{} \pi^2 \,({\ell_0\,})^{\nu-3} \,\delta(1-t)~,
\end{equation}
which explicitly counts the point $t=1$ only once per integration.
Using the Sterling approximation,
\begin{equation}
	\lim\limits_{z\rightarrow\infty} \frac{\sqrt{2\pi z}\left(z/e\right)^z}{\Gamma(z+1)} = 1~,
\end{equation}
we find that for large $\ell$ this result converges towards
\begin{equation}\label{eq_Il_limit}
I_\ell(\nu,1) \xrightarrow[\ell \to \infty]{} \pi^{\frac{3}{2}} \frac{\Gamma(1-\frac{\nu}{2})}{\Gamma(\frac{3-\nu}{2})} \frac{\ell^{\ell+\nu/2}}{\ell^{\ell+2-\nu/2}}~,
\end{equation}
which supports the statement that for large $\ell$ we have $I_\ell(\nu,1) \sim \ell^{\nu-2}$ in the Limber approximation. If we assume that the dominant $t$ dependence stems from the $t^\ell$ factor for large $\ell$ (this is a good approximation), we can immediately see that the condition $\tmin^\ell = \epsilon$ implies $\tmin=\epsilon^{1/\ell} \approx 1+ log(\epsilon)/\ell+\mathcal{O}(1/\ell^2)$, and thus $1-\tmin \approx log(1/\epsilon)/\ell+\mathcal{O}(1/\ell^2)$, which is proportional to $1/\ell$ for $\epsilon\ll1$\,. Since the support of the function shrinks like $1/\ell$ while its amplitude grows as $\ell^{\nu-2}$, we recover the argument that the Limber limit reads $\ell^{\nu-3} \delta(1-t)$ for large $\ell$\,. This is further supported by figure \ref{fig_tmin}. We have additionally confirmed using \texttt{Mathematica} and \texttt{mpmath} in Python, that the integrated area behaves like $\ell^{\nu-3}$ for large $\ell$ as expected.
\begin{figure}[h!]
	\centering
	\includegraphics[width=0.5\textwidth]{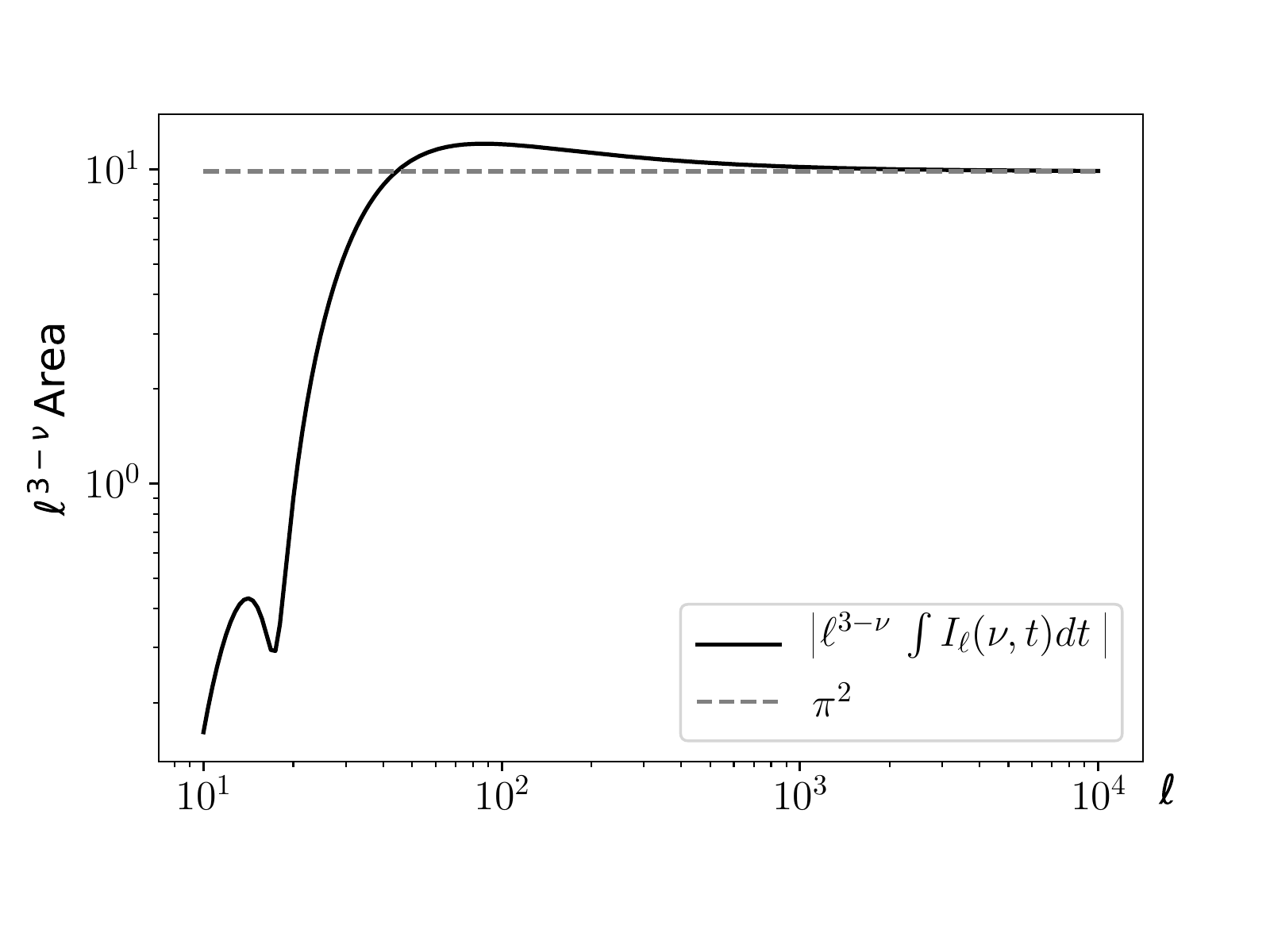}
	\caption{An illustration of the Limber limit: For large $\ell$ the area under the curve $I_\ell(\nu,t)$ approaches $\pi^2 \ell^{\nu-3}$ when integrated from $0$ to $1$. We see that the factor $\ell^{3-\nu} \, I_\ell(\nu,t)$ approaches the constant $\pi^2$, which is an equivalent statement.} \hspace*{-1em}
\end{figure}
\FloatBarrier
\noindent \textit{Transformations.} For high $t$ we can transform our original $I_\ell(\nu,t)$ using the properties of hypergeometric functions to speed up convergence, by bringing the argument closer to $0$, where the hypergeometric function's Taylor expansion converges faster.
We find for high $t$ the transformation
{\small
\begin{multline}\label{ap_eq_hight}
I_\ell(\nu,t) = \frac{\pi^{\frac{3}{2}} t^\ell \left(\frac{2}{1+z}\right)^{\ell+\nu/2}}{\Gamma\left(\frac{3-\nu}{2}\right)}\Biggl[\frac{\Gamma\left(\frac{2\ell+\nu}{2}\right)\Gamma\left(\frac{2-\nu}{2}\right)}{\Gamma\left(\frac{4+2\ell-\nu}{2}\right)} ~ {}_2F_1 \left(\frac{2\ell+\nu}{4},\frac{2\ell+\nu+2}{4},\frac{\nu}{2},\left(\frac{1-z}{1+z}\right)^2\right) 
\\+ \Gamma\left(\frac{\nu-2}{2}\right)\left(2\left(\frac{1+z}{1-z}\right)\right)^{\nu-2} {}_2F_1 \left(\frac{2\ell+6-\nu}{4},\frac{2\ell+4-\nu}{4},2-\frac{\nu}{2},\left(\frac{1-z}{1+z}\right)^2\right)\Biggr]~,
\end{multline}
}
where $z=t^2$. For small $t$, this formula depends on a precise cancellation between the two hypergeometric functions, disfavoring its use in this regime. Instead, for small $t$ we find another transformation that allows for the efficient calculation of $I_\ell(\nu,t)$
{\small
\begin{equation} \label{ap_eq_lowt}
I_\ell(\nu,t) = \frac{2^{\nu-1}\pi^2 t^\ell \left(1+z\right)^{-(\ell+\frac{\nu}{2})}}{\Gamma\left(\frac{3-\nu}{2}\right)}\frac{\Gamma\left(\frac{2\ell+\nu}{2}\right)}{\Gamma\left(\ell+3/2\right)}~ {}_2F_1 \left(\frac{2\ell+\nu}{4},\frac{2\ell+\nu+2}{4},\ell+\frac{3}{2},\frac{4z}{(1+z)^2}\right)~.
\end{equation} 
}
Notice that on the right hand side of this formula the imaginary frequencies $\nu$ in the arguments of the hypergeometric function are divided by a factor of 4. This is very helpful for faster convergence of the numerical evaluation. Similarly to \cite{Assassi:2017lea} we prescribe to switch from version (\ref{ap_eq_lowt}) to (\ref{ap_eq_hight}) above some $t^{*}$. One can easily perform numerical test to find the most optimal choice of $t^*$ is roughly $t^{*}\approx0.9$. 
\dnew
\textit{Derivatives of the Bessel Functions.} The integration by parts method in section \ref{sec_derivs} works well most of the time, but there are some situations in which it can fail: 
\begin{enumerate}
	\item When the window function is a Dirac distribution, which makes the operation of taking derivatives very imprecise.
	\item When the window function is a function similar to a top-hat, for which the second derivatives are very localized around the edges of the function. This results in narrow peaks in the $f_n^{ij}(t)$ function at values corresponding to the edge ratios.
\end{enumerate}
In our current implementation, we just discard these two cases. We can give however some hints of possible solutions. First of all, it is possible to use the following equation to translate a derivative into a change of $\ell$,
\begin{equation}
	\frac{\partial}{\partial x}\bessel{\ell}{x} = \frac{\ell}{x} \bessel{\ell}{x} - \bessel{\ell+1}{x} ~.
\end{equation}
Thus, for each derivative of the Bessel function we obtain the two integrals, $\ell I_\ell(\nu-1,t)$ and 
\begin{equation}
	4 \pi \int\limits_{0}^\infty du u^{\nu-1} \bessel{\ell+1}{u} \bessel{\ell}{u t} \equiv J_\ell^{(1)}(\nu,t) ~.
\end{equation}
Thus the integrals $I_\ell(\nu,t)$ and $J_\ell^{(1)}(\nu,t)$ provide a basis for all derivatives of Bessel functions. Sadly, the $J_\ell^{(1)}(\nu,t)$ does not have nice symmetry properties with respect to $t$ as $I_\ell(\nu,t)$ does. In order to deal only with $t<1$, we could decide to define 
\begin{equation}
	J^{(2)}_\ell(\nu,t) \equiv 4\pi \int\limits_{0}^\infty du u^{\nu-1} \bessel{\ell}{u} \bessel{\ell+1}{u t}~.
\end{equation}
and relate $J^{(1)}(\nu,t)$ for $t>1$ to $J^{(2)}(\nu,1/t)$ and vice versa.  
\dnew 
Secondly, one can relate all derivatives of the Bessel functions to the derivatives of $I_\ell(\nu,t)$, e.g. 
\begin{equation}\label{ap_eq_Ilderivs}
\begin{split}
4\pi \int\limits_{0}^\infty du u^{\nu-1} \bessel{\ell}{u} \left.\frac{\partial\bessel{\ell}{x}}{\partial x}\right|_{x=u t} & = \partial_{t}\, I_\ell(\nu-1,t)~, \\ 
4\pi \int\limits_{0}^\infty du u^{\nu-1} \left.\frac{\partial\bessel{\ell}{x}}{\partial x}\right|_{x=u} \bessel{\ell}{u t} & = -t\, \partial_{t}\, I_\ell(\nu-1,t) + (1-\nu) I_\ell(\nu-1,t)~.
\end{split}
\end{equation}
Of course, since the derivatives of the hypergeometric functions are known analytically, these expressions are also known from analytic formulas. In either case, there is a strong computational effort involved compared to the simple integration by parts method that we use for smooth window functions.
\dnew If one wants to use the relations from appendix G of \cite{Gebhardt:2017chz}, one would also have to expand the Doppler terms involving only first derivatives using the following equation
\begin{equation}
	k\bessel{\ell}{k\chi} = \frac{\ell-1}{\chi}\bessel{\ell-1}{k\chi}-\frac{\partial}{\partial \chi}\bessel{\ell-1}{k\chi}~,
\end{equation}
to keep the separation of appearing Bessel functions to $\Delta\ell=2$, which allows the given recursion relations in $\Delta\ell$ to be used. The \rsd\,term results only in $\Delta \ell=-2,0,2$\,.
\section{Further Notes}\label{sec_further}
\noindent \textit{Width measurement parameters.} We give below the precision parameters used in our tests involving different bin widths, presented in Table \ref{tab_width}.
\begin{table}[h!]
	\centering
	\begin{tabular}{| c | c c c c c c |}
		\hline
		$\Delta z$ & 0.5 & 0.1 & 0.05 & 0.01 & 0.005 & 0.01 \Tstrut{2.5}\\ \hline
		q logstep & 80 & 50 & 33 & 21 & 5 & 0.05 \Tstrut{3}\\\
		selection sampling bessel & 1.0 & 1.9 & 2.5 & 1.0 & 1.0 & 9.0\\
		hyper sampling flat & 7& 7.5 & 7 & 7 & 7 & 7\\
		\hline
	\end{tabular}
	\caption{\label{tab_width_params}Precision parameters used for the tests presented in Table \ref{tab_width}.}
\end{table}
\FloatBarrier
\noindent \textit{Cosmological parameters.} In all our tests, we used a fixed set of $\Lambda$CDM model parameters summarized in Table \ref{tab_cosmo}. The value of $\Omega_\Lambda$ is adjusted to cancel the spatial curvature. 
$Y_\mathrm{He}$ is the Helium fraction, $z_\mathrm{reio}$ is the redshift of reionization, $N_{eff}$ is the effective neutrino number. The index $\mathrm{ncdm}$ refers to non-cold Dark Matter, in our case massive neutrinos. We assumed $3$ neutrino species with degenerate mass of $0.33\mathrm{eV}$\,, for which we held constant $N_{eff}$\,, $H_0$\,, $\Omega_\Lambda$\,, and $\omega_b$\,, implying fixed $\omega_m$.
\begin{table}[h!]
	\centering
	\begin{tabular}{|| c | c | c || c | c ||}
		\hline 
		& Name & Value & Name & Value \Tstrut{2.5}\\ \hline 
		Common parameters &$h$ & $0.6711$ &$T_{cmb}$ & $2.726 \mathrm{K}$ \Tstrut{3}\\
		& $\omega_b$ & 0.02207 & $Y_\mathrm{He}$ & 0.25 \\
		& $N_{eff}$ & 3.04 & $A_s$ & $2.22 \cdot 10^{-9}$\\
		& $z_\mathrm{reio}$ & 10. & $n_s$& 0.97\\ \hline
		Massless neutrinos & $\omega_\mathrm{cdm}$ & 0.12029 & & \Tstrut{2.5}\\ \hline
		Massive neutrinos & $\omega_\mathrm{cdm}$ & 0.10965534872 & $N_\mathrm{ncdm}$& 1 \Tstrut{2.5}\\
		& $m_\mathrm{ncdm}$ & $3 \cdot 0.33 \mathrm{eV}$ & &\\ \hline
	\end{tabular}
	\caption{The cosmological parameters used in the comparison study of section \ref{sec_concl}.\label{tab_cosmo}}
\end{table}
\FloatBarrier
\noindent \textit{Discrete $\ell$ values.} The list of $\ell$ values used in section \ref{sec_concl} for testing purposes is given below.\dnew
\begin{table}[h!]
	\centering
	\begin{tabular}{c c c c c c c c c c c c}
		2&3&4&5&6&7&8&9&10&11&12&13\\14&15&16&17&19&21&23 &25&27&30&33&36\\40&44&49&54&60&67&75&83&92&103&115&128\\143 &160&179&200&223&249&278&311&348&388 &428&468\\508&548&588&628&668&708&748 &788&828&868&908&948\\988&1000 & & & & & & & & & & \\
	\end{tabular}
	\caption{The $\ell$ values for the calculation of the number count spectra ($\ell_{max}=1000$ , $N_\ell=62$).}
	\vspace*{-1em}
\end{table}
\FloatBarrier
\begin{table}[h!]
	\centering
	\begin{tabular}{c c c c c c c c c c c c}
		2&3&4&5&6&7&8&9&10&11&12&13\\14&15&16&17&19&21&23&25&27&30&33&36\\40&44&49&54&60&67&75&83&92&103&115&128\\143&160&179&200&223&249&278&311&348&388&428&468\\508&548&588&628&668&708&748&788&828&868&908&948\\988&1028&1068&1108&1148&1188&1228&1268&1308&1348&1388&1428\\1468&1508&1548&1588&1628&1668&1708&1748&1788&1828&1868&1908\\1948&1988&2028&2068&2108&2148&2188&2228&2268&2308&2348&2388\\2428&2468&2508&2548&2588&2628&2668&2708&2748&2788&2828&2868\\2908&2948&2988&3000& & & & & & & & \\
	\end{tabular}
	\caption{The $\ell$ values for the calculation of the Cosmic shear spectra ($\ell_{max}=3000$ , $N_\ell=112$).}
	\vspace*{-5em}
\end{table}
\FloatBarrier
 	\clearpage
	\bibliographystyle{JHEP}
	\bibliography{FFT_CLASS}
\end{document}